\documentclass[times,twocolumn]{aastex63}
\usepackage{amsmath}
\usepackage{amssymb}

\newcommand{\bracketfunc}[3]{\left(\frac{#1}{#2}\right)^{#3}}
\newcommand{\re}[1]{{\rm Re}\left(#1\right)}

\newcommand{\rchara}{R_{\rm c}}
\newcommand{\rhole}{R_{\rm cav}}
\newcommand{\deltahole}{\delta_{\rm cav}}
\newcommand{\ftot}{F_{\rm tot}}
\newcommand{\rhovec}{\vec{\rho}}
\newcommand{\rvec}{\vec{R}}
\newcommand{\pa}{\phi}
\newcommand{\inc}{i}
\newcommand{\rwi}[1]{W_{#1}}
\newcommand{\rhi}[1]{H_{#1}}
\newcommand{\rci}[1]{R_{\rm rc,#1}}
\newcommand{\rgin}[1]{R_{\rm g,in,#1}}
\newcommand{\rgout}[1]{R_{\rm g,out,#1}}
\newcommand{\gdeltai}[1]{\delta_{\rm gc,#1}}
\newcommand{\rhosurf}{\Sigma}
\newcommand{\msun}{M_{\odot}}

\newcommand{\hscale}{h}

\newcommand{\omegak}{\Omega_{\rm K}}
\newcommand{\gangle}{\alpha_g}
\newcommand{\slarge}{s_{\rm d,large}}
\newcommand{\vrdust}{v_{\rm R,dust}}
\newcommand{\st}{St}
\newcommand{\vk}{V_{R}}

\received{\today}
\revised{\today}
\accepted{\today}
\submitjournal{ApJ}

\shorttitle{ALMA observation of the protoplanetary disk around WW Cha}
\shortauthors{K.D. Kanagawa et al.}

\begin{document}

\title{ALMA observation of the protoplanetary disk around WW Cha: faint double-peaked ring and asymmetric structure}

\correspondingauthor{Kazuhiro D. Kanagawa}
\email{kazuhiro.kanagawa.d@vc.ibaraki.ac.jp}

\author[0000-0001-7235-2417]{Kazuhiro D. Kanagawa}
\affiliation{Research Center for the Early Universe, Graduate School of Science, The University of Tokyo, Hongo, Bunkyo-ku, Tokyo 113-0033, Japan}
\affiliation{College of Science, Ibaraki University, 2-1-1 Bunkyo, Mito, Ibaraki 310-8512, Japan}

\author[0000-0002-3053-3575]{Jun Hashimoto}
\affiliation{Astrobiology Center, National Institutes of Natural Sciences, 2-21-1 Osawa, Mitaka, Tokyo 181-8588, Japan}

\author{Takayuki Muto}
\affiliation{Division of Liberal Arts, Kogakuin University, 1-24-2 Nishi-Shinjuku, Shinjuku-ku, Tokyo 163-8677, Japan}

\author[0000-0002-6034-2892]{Takashi Tsukagoshi}
\affiliation{National Astronomical Observatory of Japan, 2-21-1 Osawa, Mitaka, Tokyo 181-8588, Japan}

\author[0000-0003-3038-364X]{Sanemichi Z. Takahashi}
\affiliation{National Astronomical Observatory of Japan, 2-21-1 Osawa, Mitaka, Tokyo 181-8588, Japan}

\author{Yasuhiro Hasegawa}
\affiliation{Jet Propulsion Laboratory, California Institute of Technology, Pasadena, CA 91109, USA}

\author[0000-0003-0114-0542]{Mihoko Konishi}
\affiliation{Faculty of Science and Technology, Oita University, 700 Dannoharu, Oita 870-1192, Japan}

\author[0000-0002-7058-7682]{Hideko Nomura}
\affiliation{National Astronomical Observatory of Japan, 2-21-1 Osawa, Mitaka, Tokyo 181-8588, Japan}

\author[0000-0003-2300-2626]{Hauyu Baobab Liu}
\affiliation{Institute of Astronomy and Astrophysics, Academia Sinica, 11F of Astronomy-Mathematics Building, AS No.1, Sec. 4, Roosevelt Rd, Taipei 10617, Taiwan, R.O.C.}

\author[0000-0001-9290-7846]{Ruobing Dong}
\affiliation{Department of Physics \& Astronomy, University of Victoria, Victoria, BC, V8P 1A1, Canada}

\author[0000-0003-4562-4119]{Akimasa Kataoka}
\affiliation{National Astronomical Observatory of Japan, 2-21-1 Osawa, Mitaka, Tokyo 181-8588, Japan}

\author[0000-0002-3001-0897]{Munetake Momose}
\affiliation{College of Science, Ibaraki University, 2-1-1 Bunkyo, Mito, Ibaraki 310-8512, Japan}

\author{Tomohiro Ono}
\affiliation{Department of Earth and Planetary Sciences, Tokyo Institute of Technology, 2-12-1 Ookayama, Meguro-ku, Tokyo 152-8551, Japan}

\author[0000-0003-1799-1755]{Michael Sitko}
\affiliation{Department of Physics, University of Cincinnati, Cincinnati, OH 45221, USA}
\affiliation{Space Science Institute, 475 Walnut Street, Suite 205, Boulder, CO 80301, USA}

\author{Michihiro Takami}
\affiliation{Institute of Astronomy and Astrophysics, Academia Sinica, 11F of Astronomy-Mathematics Building, AS No.1, Sec. 4, Roosevelt Rd, Taipei 10617, Taiwan, R.O.C.}

\author[0000-0001-8105-8113]{Kengo Tomida}
\affiliation{Astronomical Institute, Tohoku University, Sendai 980-8578, Japan}



\begin{abstract}
We present Atacama Large Millimeter/submillimeter Array (ALMA) band~6 observations of dust continuum emission of the disk around WW Cha.
The dust continuum image shows a smooth disk structure with a faint (low-contrast) dust ring, extending from $\sim 40$~au to $\sim 70$~au, not accompanied by any gap.
We constructed the simple model to fit the visibility of the observed data by using MCMC method and found that the bump (we call the ring without the gap the bump) has two peaks at 40~au and 70~au.
The residual map between the model and observation indicates asymmetric structures at the center and the outer region of the disk.
These asymmetric structures are also confirmed by model-independent analysis of the imaginary part of the visibility.
The asymmetric structure at the outer region is consistent with a spiral observed by SPHERE.
To constrain physical quantities of the disk (dust density and temperature), we carried out radiative transfer simulations.
We found that the midplane temperature around the outer peak is close to the freezeout temperature of CO on water ice ($\sim 30$~K).
The temperature around the inner peak is about $50$~K, which is close to the freezeout temperature of H$_2$S and also close to the sintering temperature of several species.
We also discuss the size distribution of the dust grains using the spectral index map obtained within the band~6 data.
\end{abstract}


\section{Introduction} \label{sec:intro}
Planets are born in a protoplanetary disk around a young star.
Recent observations have revealed substructures such as gaps, rings, and crescents in the protoplanetary disks \citep[e.g.,][]{Fukagawa2013,Akiyama2015,Akiyama2016,ALMA_HLTau2015,Momose2015,Dong2018_MWC758,Long2018,Marel2019,Soon2019,Kim2020}. 
These structure could be formed at an edge of a gap induced by disk-planet interaction \citep[e.g.,][]{paardekooper2004,Muto_Inutsuka2009b,Zhu2012,Dong_Zhu_Whitney2015,Pinilla_Ovelar_Ataiee_Benisty_Birnstiel_Dishoeck_Min2015,Kanagawa_Muto_Okuzumi_Taki_Shibaike2018}.
Alternatively, these could be associated to dust growth related to snowline \citep[e.g.,][]{Zhang_Blake_Bergin2015,Cieza2017,Macias2017,Marel_Williams_Bruderer2018,Facchini2020} and the sintering effect \citep{Okuzumi_Momose_Sirono_Kobayashi_Tanaka2016}, or secular gravitational instability \citep{Takahashi_Inutsuka2014,Takahashi_Inutsuka2016a,Tominaga_Inutsuka_Takahashi2018}.
The ring/gap structures of the dust grains also could be formed by axisymmetric gas perturbation due to evolution of luminosity of the central star \citep{Vorobyov_Elbakyan_Takami_Liu2020}.
In any case, these substructures could be reflected by planet formation and growth of dust grains which are building blocks of planets.
Direct observations of the disks help to understand how formation of the planets progresses in the disk.

Our target, WW Cha is a young star with a circumstellar disk \citep[e.g.,][]{Pascucci2016,Garufi2020} in the Chameleon I star-forming region.
The star is located at about 190~pc \citep{GaiaDR2,GaiaEDR3}.
The mass of the star is about $1M_{\odot}$, the surface temperature is 4350~K (Spectral type is K5) \citep{Luhman2007},  and the luminosity is $11L_{\odot}$ \citep{Garufi2020}.
The star is very young ($\sim 0.2$~Myr) and it could be still embedded into the molecular cloud core with a high extinction \citep{Ribas2013,Garufi2020}.
A high accretion rate onto the star, $10^{-6.6} \msun/\mbox{yr}$, is inferred from the photometric and the Balmer continuum observations \citep{Manara2016}.
Moreover, the binary with the separation of $\sim 1$~au is reported by VLTI \citep{Anthonioz2015}.
The disk of WW Cha may be a pre-transition disk because strong infrared emission is detected \citep{Espaillat2011,Ribas2013}, while the recent modeling using radiative transfer simulations done by \cite{Marel2016} suggested an inner cavity with the radius of $\sim 50$~au (but with a large uncertainty).
The disk is very bright in millimeter wavelength \citep{Pascucci2016} and \cite{Lommen2009} reported the emission at $\sim 1$~cm, which indicates the presence of large grains due to growth of dust grains.

In this paper, we report dust continuum observations of the disk around WW Cha in ALMA Cycle 5.
In Section~\ref{sec:obs}, we describe the setup of the observation and show the observational results.
We developed a model for the observed emission by using the Markov Chain Monte Carlo (MCMC) method and found axisymmetric substructures and asymmetric structures, which is described in Section~\ref{sec:modeling}.
Moreover, we carried out radiative transfer simulations to constrain the physical parameters of the disk which are described in Section~\ref{sec:rt}.
In Section~\ref{sec:discussion}, we discuss origins of the substructures, an inner cavity and binary, and the dust growth in the disk.
Section~\ref{sec:conclusion} contains our conclusion.

\section{Observations and results} \label{sec:obs}
The observation was carried out by ALMA in band 6, which is summarized in Table~\ref{tab:obs}.
The data were calibrated by the Common Astronomy Software Applications (CASA) package \citep{McMullin2007} version 5.6.1-8, following the calibration scripts provided by ALMA.
We conducted self-calibration of the visibilities. 
The phases were self-calibrated once with a fairly long solution intervals (solint=`inf') combining all spectral windows (SPWs). 

We combined the data taken by sparse (C43-8) and compact (C43-4) array configurations to recover the missing flux at larger angular scales.
By using the CASA tool \verb#uvmodelfit#, we fitted the data by a Gaussian shape, and the phase center was corrected to be the center of the Gaussian shape  by \verb#fixvis#.
The inclination $\inc=37.2^\circ \pm 0.026^{\circ}$ and position angle $\pa=32.4^\circ \pm 0.04^{\circ}$ are obtained by the Gaussian fit by \verb#uvmodelfit# \footnote{The fitted inclination and position angle are slightly different in the C43-8 (sparse configuration) and C43-4 (compact configuration) data. We adopted the values of the C43-8 data. For the C43-4 data, $\inc=39.2^{\circ}$ and $\pa=30.9^{\circ}$ with the relatively large reduced $\chi^2$, 5.78 while, for the fit of the C43-8 data, reduced $\chi^2=1.89$. The relatively large $\chi^2$ for the C43-4 data can be due to the asymmetric structure discussed in Section~\ref{subsec:asymmetricity}. Hence, we adopted the values given by the fit of the C43-8 data}.
There are two SPWs for continuum with $1.875$~GHz frequency width with the central frequency being $233.0$~GHz (Upper band) and $216.7$~GHz (Lower band) in both the C43-4 and C43-8 data.
As shown below, the total flux density of the data in $233.0$~GHz ($\sim 500$~mJy) is significantly larger than that in $216.7$~GHz ($\sim 430$~mJy).
Hence, the dust continuum image of combined data was synthesized by CASA with the \verb#tclean# task using the \verb#mtmfs# algorithm \citep{rau11} with \verb#nterms#=2.
We obtained a synthesized image at $224.9$~GHz with the beam size of $89.6\times 60.0\mbox{ mas}$ ($17.0 \times 11.4$~au) with PA=$168.8^{\circ}$ and with the $1\sigma$ RMS noise level of $0.029\mbox{ mJy/Beam}$.
The imaging parameters are summarized in Table~\ref{tab:obs}.

\begin{deluxetable}{lcc}
\tabletypesize{\footnotesize}
\tablewidth{0pt} 
\tablenum{1}
\tablecaption{ALMA band 6 Observations and Imaging Parameters\label{tab:obs}}
\tablehead{
\colhead{Observations} & \colhead{Sparse configuration} & \colhead{Compact configuration}
}
\startdata
Observing date (UT)         & 2017.Nov.27 		& 2018.Mar.11 \\
Configuration               & C43-8 			& C43-4 \\
Project code                & \multicolumn{2}{c}{2017.1.00286.S} \\
Time on source (min)        & 60.3 			& 29.4  \\
Number of antennas          & 47 			& 42    \\
Baseline lengths            & 92.1~m to 8.5~km 		& 15.1~m to 1.2~km  \\
Baseband Freqs. (GHz)       & \multicolumn{2}{c}{233.0 (Upper band), 216.7 (Lower band)} \\
Channel width   (GHz)       & \multicolumn{2}{c}{1.87} \\
Continuum band width (GHz)  & \multicolumn{2}{c}{4.0} \\
Bandpass calibrator         & J0635$-$7516		&J1427$-$4206 \\
Flux calibrator             & J0635$-$7516		&J1427$-$4206 \\
Phase calibrator            & J1058$-$8003		&J1058$-$8003 \\
Mean PWV (mm)               & 1.5 			&0.6 \\
\hline \hline
\multicolumn{1}{c}{Imaging} & \multicolumn{2}{c}{} \\
\hline
Robust clean parameter      & \multicolumn{2}{c}{0.0}\\
Deconvolution algorithm     & \multicolumn{2}{c}{mtmfs}\\
Weighting                   & \multicolumn{2}{c}{Briggs}\\
nterms						& \multicolumn{2}{c}{2}\\
Beam shape                  & \multicolumn{2}{c}{89.6~$\times$~60.0~mas ($17.0 \times 11.4$~au) at PA of $168.8^{\circ}$}\\
r.m.s. noise (mJy/beam) & \multicolumn{2}{c}{0.029} 
\enddata
\end{deluxetable}

\begin{figure*}
\gridline{\fig{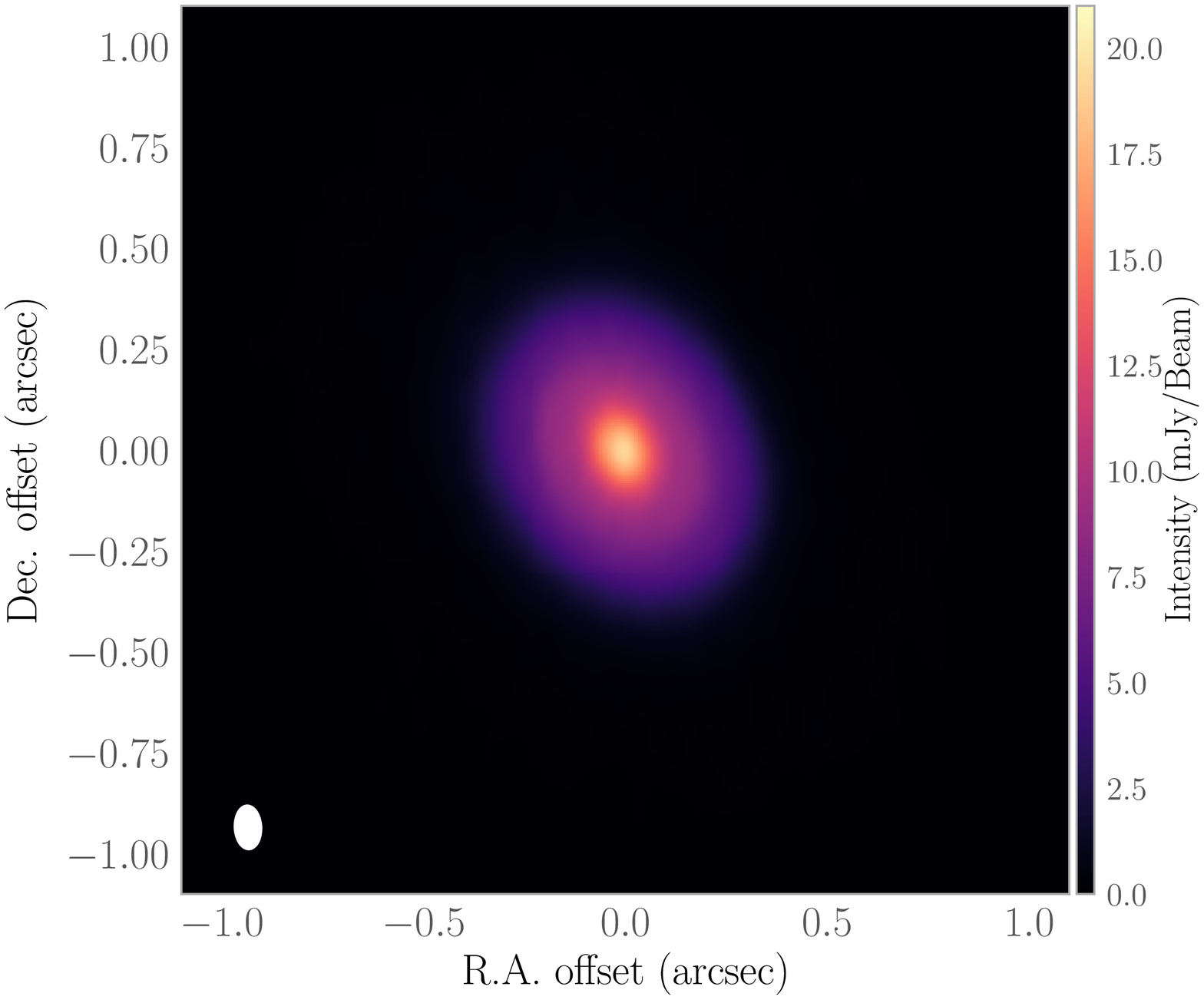}{0.49\textwidth}{(a) Combined image at $224.9$~GHz}
\fig{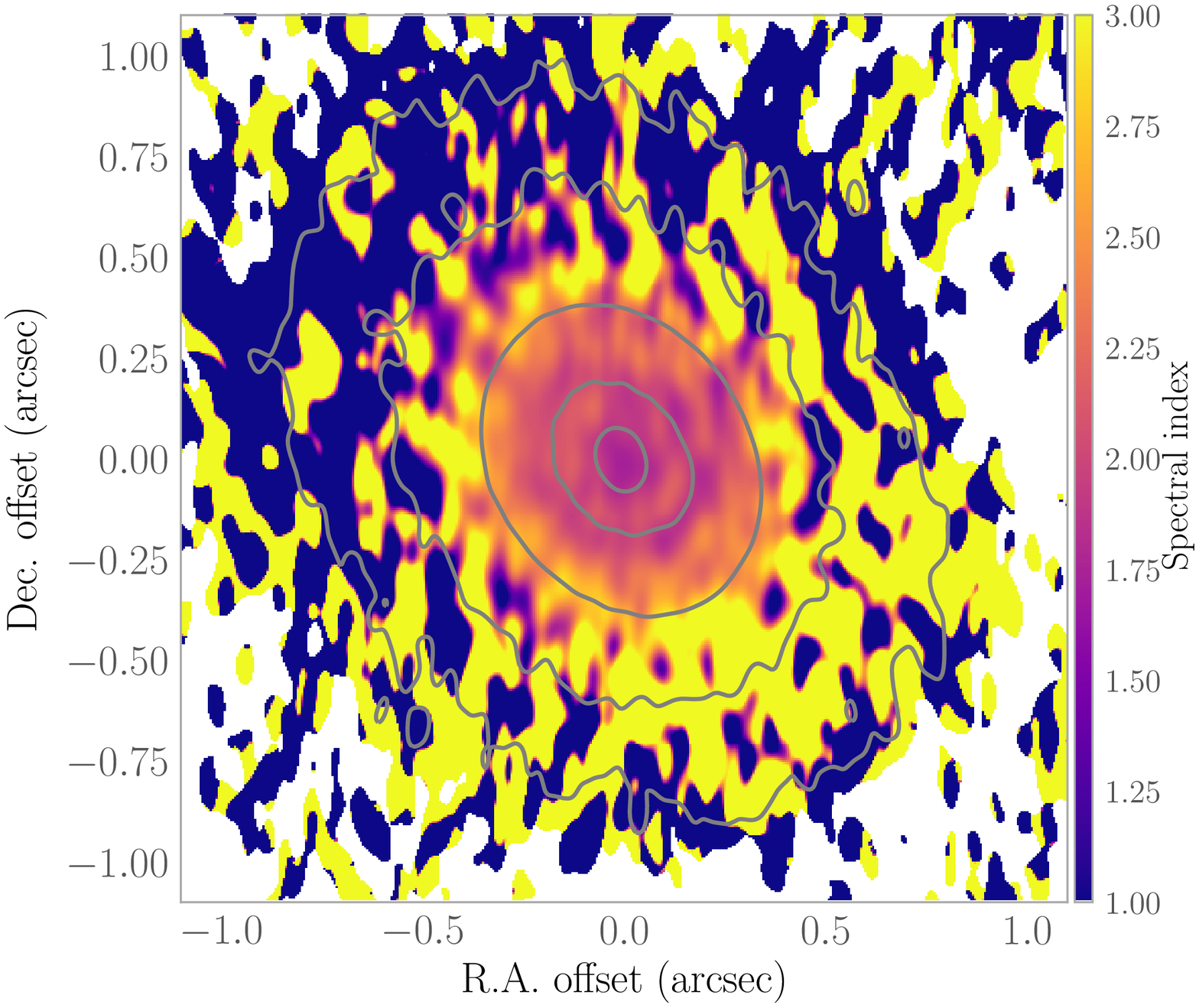}{0.49\textwidth}{(c) Image of spectral index}
}
\gridline{\fig{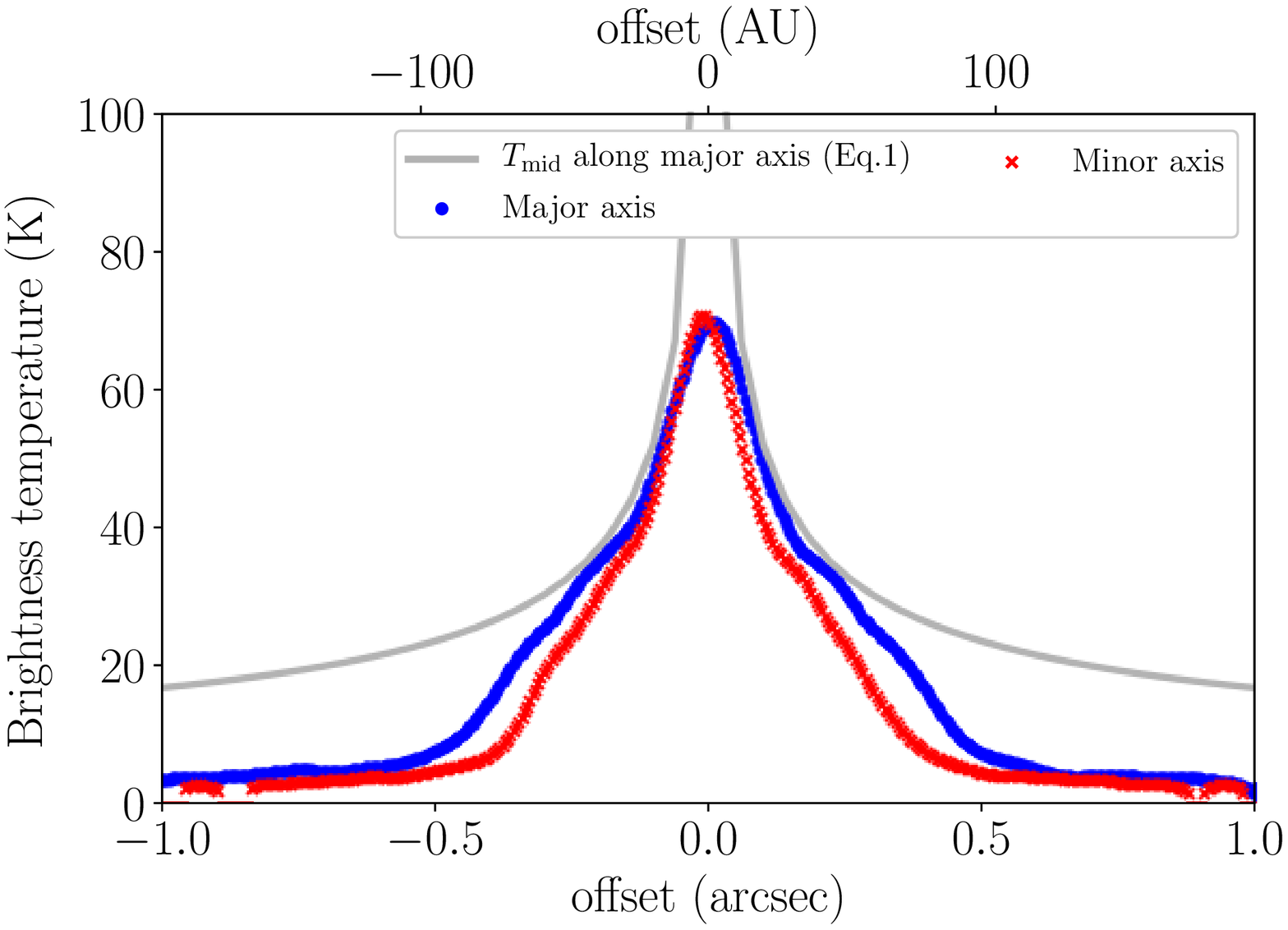}{0.49\textwidth}{(b) Radial slices of Panel (a)}
\fig{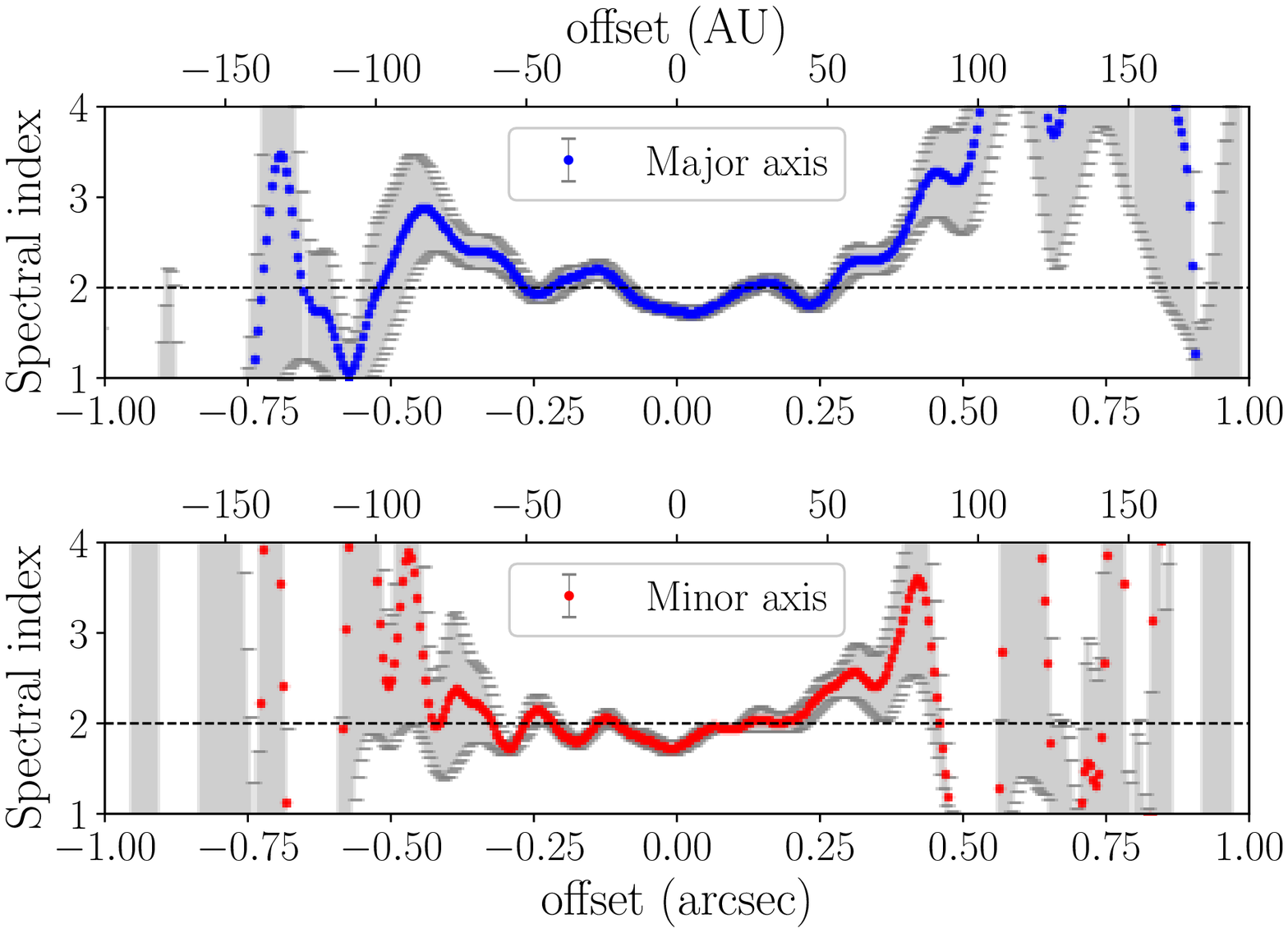}{0.49\textwidth}{(d) Radial slice of Panel (c)}
}
\caption{
panels (a) and (c) show the image and the spectral index maps resulting from the combination of 233.0~GHz and 216.7GHz data. 
The contours in Panel (c) indicate the intensity levels of $0.087$~mJy/Beam (3$\sigma$), $0.29$~mJy/Beam (10$\sigma$), $2.9$~mJy/Beam ($100\sigma$), $8.7$~mJy/Beam ($300\sigma$) and $14.5$~mJy/Beam ($500\sigma$) in Panel (a). 
Panels (b) and (d) show brightness temperature and spectral index along the major and minor axis, respectively.
The gray thick lines in Panel (b) denote the midplane temperature given by Equation~(\ref{eq:tmid}) with $L_{\ast}=11L_{\odot}$.
\label{fig:obs}}
\end{figure*}
The synthesized image is shown in Figure~\ref{fig:obs}.
The panel (a) shows the synthesized dust continuum image derived from all SPW data, and the brightness temperatures along the major and minor axes are shown in the panel (b).
The dust disk is clearly resolved and its size is about $0.5$~arcsec, which corresponds to about $100$~au.
We see a faint low-contrast dust ring feature in the radial profile of brightness temperature at $\sim 0.3$~arcsec ($\sim 60$~au) from the central star.
This ring structure is not accompanied with the gap structure, as different from the rings found in the disk of HL~Tau \citep{ALMA_HLTau2015}, and DSHARP's samples \citep{DSHARP1}.
Hence, we call this ring structure (without the gap) the bump in the following.
The total flux density with $> 3\sigma (0.087\mbox{ mJy})$ emission is measured to be $449.58$~mJy.
We do not see a clear cavity structure in the image.
Therefore, the large cavity such as that predicted by SED analysis done by \cite{Marel2016}, is ruled out at least in the millimeter image, while the existence of a cavity that is smaller than the beam size is not ruled out.

In the panel (c) of Figure~\ref{fig:obs}, we show the spectral index map, and panel (d) illustrates the spectral indexes along the major and minor axes.
Within the region of $<0.25$~arcsec ($<50$~au), the spectral index is $\sim 2$, which is an indicative of optically thick dust emission.
Around the center of the disk, in particular, the spectral index is slightly below $2$, which may indicate optically thick dust scattering \citep{Liu2019,Zhu2019}.
In the region where the offset is larger than $0.25$~arcsec, the disk may be optically thin because the spectral index is larger than 2, and the spectral index seems to increase in the outer region, though there is large uncertainly at $>0.5$~arcsec.

In the panel (b) of Figure~\ref{fig:obs}, we also plot the midplane temperature along the major axis, estimated by the simple expression for a passive heated radiative disk \citep[e.g.,][]{Chaing_Goldreich1997},
\begin{align}
T_{\rm mid} &= \left( \frac{\gangle L_{\ast}}{8\pi R^2 \sigma_{\rm SB}}\right)^{1/4},
\label{eq:tmid}
\end{align}
where $\sigma_{\rm SB}$ is the Stefan-Bolzmann constant, and $R$ is the distance from the star.
The grazing angle $\gangle$ is set to be $0.02$ and the stellar luminosity $L_{\ast}$ is $11L_{\odot}$ in the plot.
The brightness temperature is close to the midplane temperature within $0.25$~arcsec from the center, which indicate the optically thick emission.
The outer region ($R>0.25$~arcsec or $ > 50$~au) can be optically thin, which is consistent with the spectral index map mentioned above.

\begin{figure}
\plotone{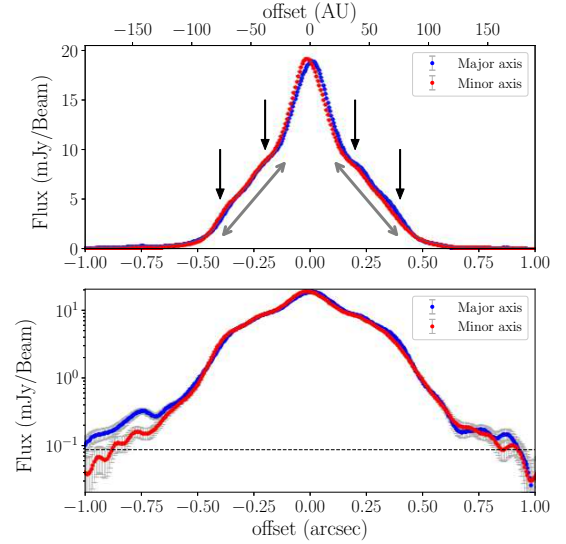}
\caption{
Intensity distributions along the major and minor axes of the face-on view of Figure~\ref{fig:obs}~(a).
We also plot the 1$\sigma$ noise level at each data point. 
The black arrows indicate the locations of the peaks within the bump structure, and the gray one indicates location of the bump structure.
The dashed horizontal line denotes $3\sigma$ noise level (=$0.087$~mJy/Beam).
The bottom panel is the same as the top panel, but the vertical axis is in logarithmic scale.
The gray double-sided arrows indicate the location of the faint dust bump and the black arrows denote the locations of the peaks within the bump.
\label{fig:obs_slice}}
\end{figure}
\begin{figure}
\plotone{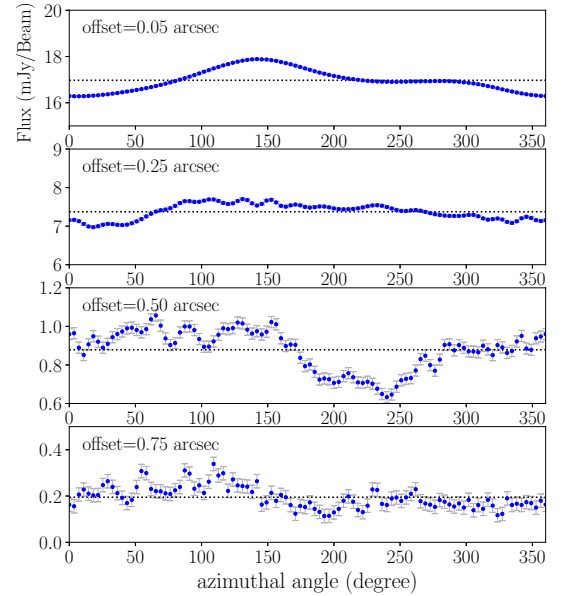}
\caption{
Azimuthal distributions of observed intensity in the face-on view, at offset $=0.05$~arcsec ($\simeq 10$~au), $0.25$~arcsec ($\simeq 50$~au), $0.5$~arcsec ($\simeq 100$~au), $0.75$~arcsec ($\simeq 150$~au).
The error bars indicate $1\sigma$ noise level (0.029~mJy).
The horizontal lines are the averages, which indicates $16.98$~mJy/Beam ($\sim 600 \sigma$), $7.37$~mJy/Beam ($\sim 250\sigma$), $0.88$~mJy/Beam ($\sim 30\sigma$), and $0.20$~mJy/Beam ($\sim 7\sigma$), from the top panel to bottom panel.
}
\label{fig:azimdist}
\end{figure}
Using the inclination and position angle, we have deprojected the face-on equivalent view, to identify substructures on the disk.
In Figure~\ref{fig:obs_slice}, we show the intensity profile along the major axis and minor axis in the face-on view.
In Figure~\ref{fig:obs_slice}, we put gray double-sided arrows to indicate the location of the faint low-contrast dust bump, which extends from $R\simeq 0.2$~arcsec ($\simeq 40$~au) to $R\simeq 0.4$~arcsec ($\simeq 80$~au).
Moreover, one can see this bump has a double-peak feature, which is indicated by the black arrows in the figure: the inner peak locates at $|R|\simeq 0.2$~arcsec ($40$~au) and the outer one locates at $|R|\simeq 0.35$~arcsec ($70$~au), where $R$ is an offset from the center.
Although the double-peak feature is not clear in the image, it is confirmed by the visibility fitting described in Section~\ref{sec:modeling}.
Outside of $0.5$~arcsec, the intensity decreases quickly, but around $|R|\sim 0.75$~arcsec ($R\sim 150$~au), the slope of the intensity becomes moderate.

Figure~\ref{fig:azimdist} shows the azimuthal distributions of the intensity in the face-on view.
There are asymmetric structures in the azimuthal distribution at $R=0.05$~arcsec ($\sim 10$~au), and the deviation from the averaged value of the averaged value at this radius is at most about 1~mJy/Beam (5\% of the averaged value or $\sim 30\sigma$).
The asymmetry at the innermost radii of the disk is also indicated in Figure~\ref{fig:obs_slice} as the distribution along the major and minor axes do not overlap at $R<20$~mas or $<4$~au.
At $R=0.25$~arcsec ($\sim 50$~au), the structure has the similar pattern of asymmetricity seen in the distribution at $R=0.05$~arcsec, and the derivation from the averaged value is about $0.3$~mJy/Beam (it is about 5\% of the averaged value or $\sim 10\sigma$).
At $R=0.5$~arcsec ($\sim 100$~au), one can see a significant asymmetric structure.
The intensity at $< 150^{\circ}$ is larger than averaged value by $0.18$~mJy/Beam ($\sim 6\sigma$) and it is smaller around $250^{\circ}$ by $0.25$~mJy/Beam ($\sim 9\sigma$).
The deviation from the averaged value is about 20\% of the averaged value.
At $R=0.75$~arcsec ($\sim 150$~au), we may see an asymmetric structure, as the intensity at $\sim 100^{\circ}$ is larger in $0.14$~mJy/Beam ($\sim 5\sigma$) than the average and it is smaller than the average at $>200^{\circ}$ in $0.08$~mJy/Beam ($\sim 3\sigma$).
We discuss asymmetric structures in Section~\ref{subsec:asymmetricity} in more detail by directly analyzing the data in visibility domain.
 
\begin{figure}
\plotone{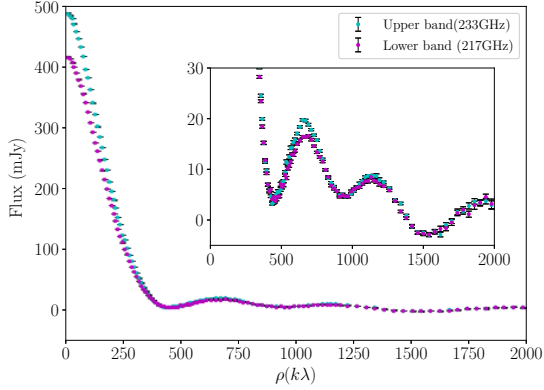}
\caption{
Real part of the visibility for the upper and lower band data.
In the plot, we combine the C43-4 and C43-8 data.
The inset shows the zoom in of the region with the flux $<30$~mJy.
\label{fig:vis_diffband}
}
\end{figure}
In the rest of this section, we show the difference between the upper ($233.0$~GHz) and lower ($216.7$~GHz) band data.
In Figure~\ref{fig:vis_diffband}, we compare the real parts of the visibility of upper and lower band data.
The visibility data are deprojected using the inclination and position angle derived earlier.
Then, the data with similar $uv$-distance are binned and averaged.
The bin width are $10k\lambda$ for $\rho<1200k\lambda$ while $20k\lambda$ for $\rho>1200k\lambda$, where $\rho$ is the $uv$-distance of the deprojected visibility data.
The error bars in the figure indicate the standard deviation of the data divided by the square root of the number of data. 
The amplitudes of the visibility are clearly different at small $uv$-distances, namely $\rho<200k\lambda$, which corresponds to a spacial scale of $\sim 1$~arcsec.
Moreover, one can find that the visibility of the upper band data is larger than that of the lower band data around the peak around $\rho=700k\lambda$.
The details of statistics of the data are described in Appendix~\ref{sec:vis_stat}.
 
\begin{figure}
\plotone{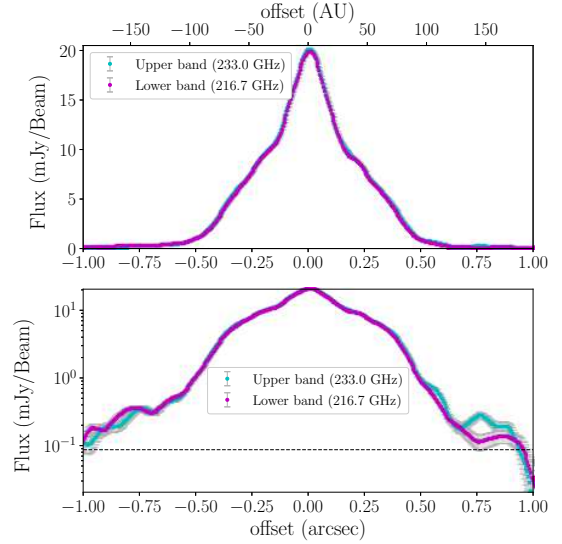}
\caption{
Intensity distributions along the major axis for the upper and lower band data.
The error bar denotes $1\sigma$ noise level (=$0.029$~mJy/Beam) and the dashed horizontal line denotes $3\sigma$ noise level ($0.087$~mJy/Beam). 
\label{fig:flux_diffband}}
\end{figure}
Figure~\ref{fig:flux_diffband} compares the intensity distributions along the major axis for the upper and lower band data.
The intensity of the upper band data is slightly larger than that of the lower band data.
Indeed, for the upper band data, the total flux density with $>3\sigma$ ($0.087$~mJy) emission is $516.16$~mJy and for the lower band data, it is $430.20$~mJy.
We note that at the offset $\simeq 0.75$~arcsec, the profile is different in the upper and lower band data, which can be related to the asymmetric structures discussed in Section~\ref{subsec:asymmetricity}.
 
\section{MCMC modeling of dust continuum emission} \label{sec:modeling}
\subsection{Model description} \label{subsec:model_description} 
To examine the structure of the disk in detail, we performed fitting for dust continuum emission in the visibility domain with a simple disk model.
As described in the previous section, the disk could have two peaks at $\sim 0.3$~arcsec and $0.75$~arcsec.
Moreover, we also include an unresolved small cavity.
By motivated from the observed features, we adopted a simple power-law intensity profile with exponential cutoff with two Gaussian bumps, two intensity enhanced/depleted regions and an inner cavity, as described below:
\begin{align}
I(R) &\propto \left[ f_{I}\bracketfunc{R}{\rchara}{-\gamma} \exp\left[-\bracketfunc{R}{\rchara}{\zeta}\right]  \right. \nonumber \\
& \qquad \qquad + \left. \sum_{i=1}^{2} \rhi{i} \exp \left[-\bracketfunc{R - \rci{i}}{\rwi{i}}{2} \right] \right],
\label{eq:I_dist}
\end{align}
where
\begin{align}
f_I &= \begin{cases}
\deltahole & 0 < R < \rhole \\
1 & \rhole < R < \rgin{1} \\
\gdeltai{1} & \rgin{1} < R < \rgout{1}) \\
\gdeltai{2} & R > \rgout{1}) \\
\end{cases}
\label{eq:f}
\end{align}
The total flux that the intensity given by Equation~(\ref{eq:I_dist}) is integrated over the entire disk is normalized to be $\ftot$ which is one of the parameters of the model.
The intensity distribution of Equation~(\ref{eq:I_dist}) has 16 parameters, namely, two exponents $\gamma$ $\zeta$, the depth and radius of the inner cavity $\deltahole,\rhole$, characteristic radius $\rchara$, total intensity of the disk $\ftot$, and parameters of substructures: $\rgin{1},\rgout{1},\gdeltai{1},\gdeltai{2}$ for two enhanced/depleted regions and $\rci{1},\rwi{1},\rhi{1}$,$\rci{2},\rwi{2},\rhi{2}$ for two Gaussian bumps.

\subsection{Fitting approach} \label{subsec:mcmc}
We fit the observation data with the model of Equation~(\ref{eq:I_dist}) in the visibility domain. 
In this modeling, we focus on a symmetric structure, following \cite{Zhang_Blake_Bergin2015}.
In the following, $\rho$ indicates the deprojected baseline in the visibility domain.
The likelihood function is defined by
\begin{align}
\chi^2 &= \sum_{k}^{N} \left(\frac{\re{\overline{V}}_{\rm obs,k} - \re{\overline{V}}_{\rm model,k}}{\sigma_{\rm obs,k}} \right)^2,
\label{eq:chi2}
\end{align}
where $k$ indicates the index of the radial bin and $N$ is the total number of radial bins.
We take an average within the bin in radial and azimuthal direction in visibility domain (the overline indicates the average).
The bin size is $10k\lambda$ for $\rho>1200k\lambda$, and $20k\lambda$ for $\rho<1200k\lambda$.
Since the amplitude of the visibility is comparable with the noise level in $\rho>2000k\lambda$, we used the visibility in the range of $\rho < 2000k\lambda$ in this modeling.
The real part of the visibility is denoted by $\re{V}$ and the subscript 'model' and 'obs' indicate the quantities of model and observation, respectively.
The standard deviation of the averaged real part of the visibility $\sigma_{\rm obs,i}$ is calculated by dividing the standard deviation of azimuthal direction in the visibility domain by the square root of the number of data within the bin.

For the fitting, we utilized the public python code \verb#vis_sample# \citep{Loomis2017}.
We used the Markov Chain Monte Carlo (MCMC) method in the \verb#emcee# package \citep{emcee}.
We carried out the fitting with the MCMC method with $\chi^2$ given by Equations~(\ref{eq:chi2}).
In the MCMC fitting, we run 1000 steps with 100 walkers after the burnin phase with 1000 steps (2000 steps in total).

\subsection{fitting result} \label{subsec:mcmc_result}
We found that the C43-4 (compact configuration) data is slightly scattered as compared with the C43-8 (sparse configuration) data, and the data is slightly statistically different around $\rho=200k\lambda$ (see Appendix~\ref{sec:vis_stat}).
Because of this difference between the C43-4 and C43-8 data, the reduced $\chi^2$ of the best fit model is much deviated from unity, when the data are combined.
If only C43-4 data is used, the resolution is not enough to identify substructures. 
Hence, we used only the C43-8 data for the MCMC fitting \footnote{When both the C43-4 and C43-8 data are used, the reduced $\chi^2$ is $\sim 6$, though the best fit parameters are similar to these shown in table~\ref{tab:bestfit}. This large reduced $\chi^2$ is mainly due to points around $\rho = 200k\lambda$.}.
We performed the MCMC fitting for the upper and lower band data separately.

The total flux density of the images synthesized by only the C43-8 data is $10\%$ -- $20\%$ smaller than that by both C43-8 and C43-4 data due to missing flux.
However, since the visibilities of the C43-8 data are quite similar to these that combined by the C43-8 and C43-4 data, as shown in Figure~\ref{fig:vis_real_spws}), excluding the C43-4 data could not affect the fitting results.

\begin{deluxetable*}{c|ccc|ccc}
\tabletypesize{\footnotesize}
\tablewidth{0pt} 
\tablenum{2}
\tablecaption{Fitting results for C43-8 data \label{tab:bestfit}}
\tablehead{
\colhead{} \vline& \multicolumn{3}{c}{Upper band (233.0~GHz data)} \vline&  \multicolumn{3}{c}{Lower band (216.7~GHz data)}\\
\colhead{Parameters} \vline& \colhead{Best fit} & \multicolumn{2}{c}{Range} \vline&\colhead{Best fit} & \multicolumn{2}{c}{Range} \\
\colhead{}\vline& \colhead{}& \colhead{Min} & \colhead{Max} \vline& \colhead{}& \colhead{Min} & \colhead{Max}
}
\startdata
$\gamma$ 			& $ 0.349^{-0.025}_{+0.030}$	& 0.0 	& 0.5	& $ 0.280^{-0.021}_{+0.023}$	& 0.0 	& 0.5\\
$\zeta$ 			& $ 1.946^{-0.061}_{+0.079}$	& 1.8	& 2.3	& $ 1.489^{-0.041}_{+0.062}$	& 1.0 	& 2.0\\
$R_{\rm c}$ (au)	& $ 51.160^{-1.777}_{+2.249}$	& 38.0 	& 57.0	& $ 45.881^{-1.499}_{+1.710}$	& 26.6 	& 76.0\\
$R_{\rm cav}$ (au)	& $ 1.296^{-0.490}_{+0.348}$	& 0.0 	& 1.9	& $ 0.986^{-0.450}_{+0.353}$	& 0.0 	& 1.9\\
$R_{\rm g,in,1}$ (au) & $ 35.908^{-0.211}_{+0.216}$	& 30.4 	& 45.6	& $ 35.630^{-0.314}_{+0.245}$	& 30.4 	& 45.6\\
$R_{\rm g,out,1}$ (au) & $ 81.994^{-2.973}_{+3.488}$	& 57.0 	& 114.0	& $ 71.894^{-1.949}_{+1.804}$	& 57.0 	& 114.0\\
$R_{\rm rc,1}$ (au) & $ 67.046^{-0.240}_{+0.216}$	& 57.0 	& 76.0	& $ 68.310^{-0.742}_{+0.775}$	& 57.0 	& 76.0\\
$W_{\rm 1}$ (au)  & $ 10.835^{-0.553}_{+0.613}$	& 0.4 	& 38.0	& $ 10.990^{-0.621}_{+1.308}$	& 0.4 	& 38.0\\
$\ln(H_{\rm 1})$ 		& $ -0.627^{-0.019}_{+0.026}$	& -1.0 	& 0.5	& $ -0.826^{-0.024}_{+0.036}$	&-1.0	& 0.5\\
$R_{\rm rc,2}$ (au) & $ 122.223^{-6.281}_{+5.861}$	& 95.0 	& 152.0	& $ 146.630^{-6.825}_{+4.100}$	& 114.0 & 190.0\\
$W_{\rm 2}$ (au)	& $ 56.530^{-3.527}_{+3.597}$	& 11.4 	& 76.0	& $ 42.890^{-3.547}_{+5.221}$	& 11.4 	& 76.0\\
$\ln(H_{\rm 2})$ 	& $ -1.708^{-0.031}_{+0.051}$	& -3.0 	& -1.0	& $ -2.036^{-0.050}_{+0.060}$	&-3.0 	& -1.5\\
$\ln(\delta_{\rm cav})$ 	& $ -1.055^{-0.763}_{+0.498}$	& -3.0 	& -0.0	& $ -1.517^{-0.845}_{+0.781}$	&-3.0 	& -0.0\\
$\ln(\delta_{\rm g,1})$ 	& $ 0.194^{-0.010}_{+0.009}$	& -0.5 	& 1.0	& $ 0.199^{-0.010}_{+0.008}$	& -0.5 	& 1.0\\
$\ln(\delta_{\rm g,2})$ 	& $ 0.305^{-0.045}_{+0.037}$	& -1.0 	& 1.0	& $ 0.091^{-0.063}_{+0.055}$	& -1.0 	& 0.5\\
$F_{\rm tot}$ (mJy) 		& $ 493.995^{-0.553}_{+0.527}$	& 470.0	& 510.0	& $ 418.431^{-0.593}_{+0.558}$	& 390.0 & 430.0\\
\enddata
\tablecomments{
Error range of the best parameters are estimated by $\pm 1 \sigma$.
}
\end{deluxetable*}
The fitting results are summarized in Table~\ref{tab:bestfit}.
The best fit parameters for the upper and lower band data are slightly different, especially on the total flux density, and the parameters related the outer structure, namely, $\zeta$, $\delta_{g,2}$ and the parameters of the outer peak ($\rci{2},\rwi{2},\rhi{2}$), and the depth of the inner cavity ($\delta_{\rm cav}$).

\begin{figure}
\plotone{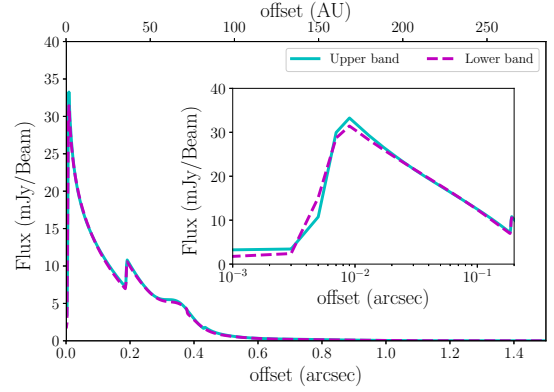}
\caption{
The shapes of the best-fit model given by parameters listed in Table~\ref{tab:bestfit}, for the upper and lower band data.
The inset is the zoom in of the region of offset $< 0.2$~arcsec.
\label{fig:shape_bestfit}}
\end{figure}
Figure~\ref{fig:shape_bestfit} shows the best-fit models for the upper and lower band data.
They look very similar, as both have small cavity with radius $\sim 1$~au (corresponding to $R_{\rm cav}$), a bump-like excess emission at $R\sim 0.2$ -- $0.4$~arcsec with double peaks at $0.2$~arcsec (corresponding to $\rgin{1}$, $\simeq 40$~au) and $0.35$~arcsec (corresponding to $\rci{1}$, $\simeq 70$~au).
The fitting results of both upper and lower bands indicate that the there is a cavity with the radius of $\sim 0.005$~arcsec ($\sim 1$~au).
Since the size of inner cavity is much smaller than the spatial resolution with $\rho<2000 k\lambda$, we consider that this structure should be confirmed in future observations.
For the outer region, though the parameters such as $\gamma$, $\zeta$, and $R_{\rm rc,2}$ are slightly different, the profiles agree with each other.

\begin{figure}
\plotone{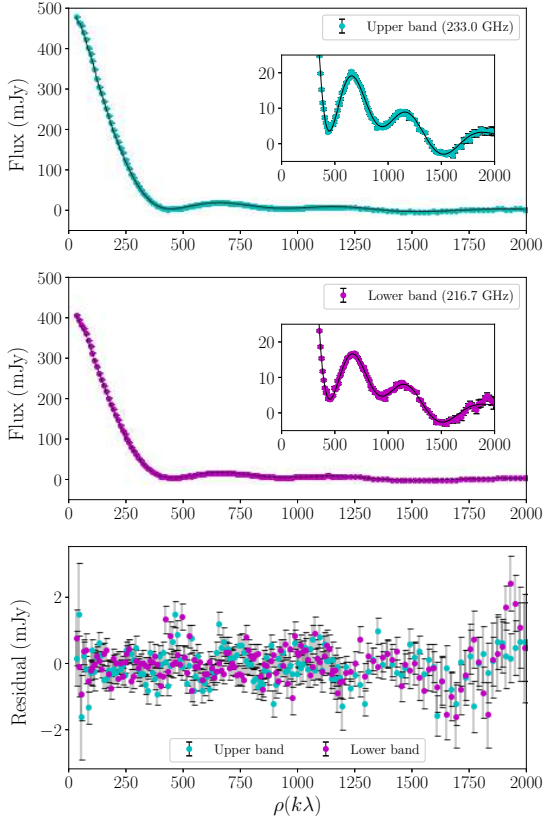}
\caption{
Real part of the visibilities for the observation (dot) and best-fit-model (solid)) for the upper and lower band data, in the upper and middle panels, respectively.
The inset shows the zoom-up view of the region with $<30$~mJy.
In the lower panel, we show the different of the visibilities between the observation and the model.
The error bars of the observational data and residuals are estimated by $1\sigma$ deviation of the average.
\label{fig:comp_vis}}
\end{figure}
Figure~\ref{fig:comp_vis} compares the visibility of the model and observation.
As can be seen in the figure, the models well reproduce the observed visibility, and the reduced $\chi^2$ for the model of the upper band data is $1.58$ and that of the lower band data is $1.28$, respectively.

\begin{figure}
\plotone{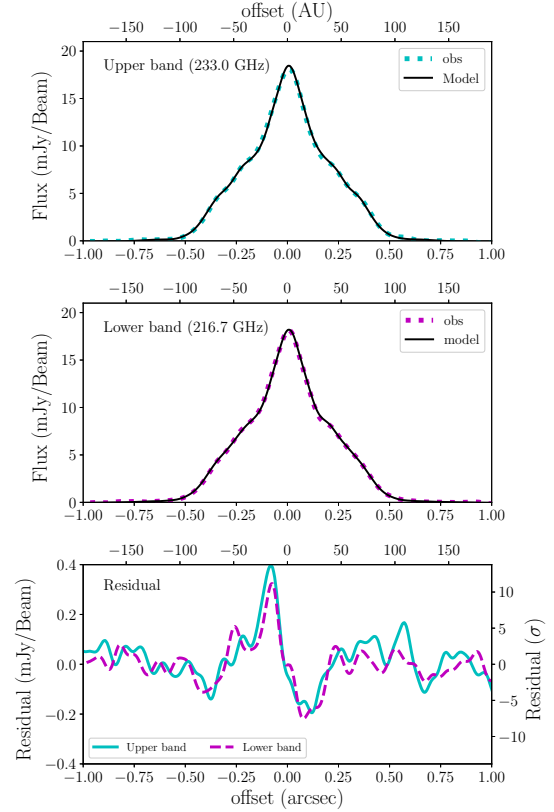}
\caption{
Intensity distributions along the major axis in the models and observations for the upper band data(upper) and the lower band data (middle).
The bottom panel shows the residuals between the model and the observation, along the major axis.
\label{fig:comp_flux}}
\end{figure}
Figure~\ref{fig:comp_flux} compares the model and the observation in the image.
The model image is first converted to the ALMA measurement set by \verb#vis_simple# with the observed measurement set and we made a mock observational image by the same procedure of the imaging of the observed data with the parameters listed in Table~\ref{fig:obs}.
In Figure~\ref{fig:comp_flux}, we show the intensity distributions along the major axis in the mock observational image (model) and the observed image.
In the bottom panel of the figure, we show the residual between the model and the observational data.
We calculated the residual as the subtraction between the model and the observation in the visibility domain, by using \verb#vis_sample#.
The visibility of the residual is converted by \verb#tclean# task with the imaging parameters listed in Table~\ref{tab:obs}.
Around the center, one can see the residual which is larger than $3\sigma$, though the residual is smaller or comparable with the $3\sigma$ level in other regions.
This discrepancy is related to the asymmetry which is indicated by the difference between the structures along the major and minor axes shown in Figure~\ref{fig:obs_slice}.

\begin{figure*}
\gridline{\fig{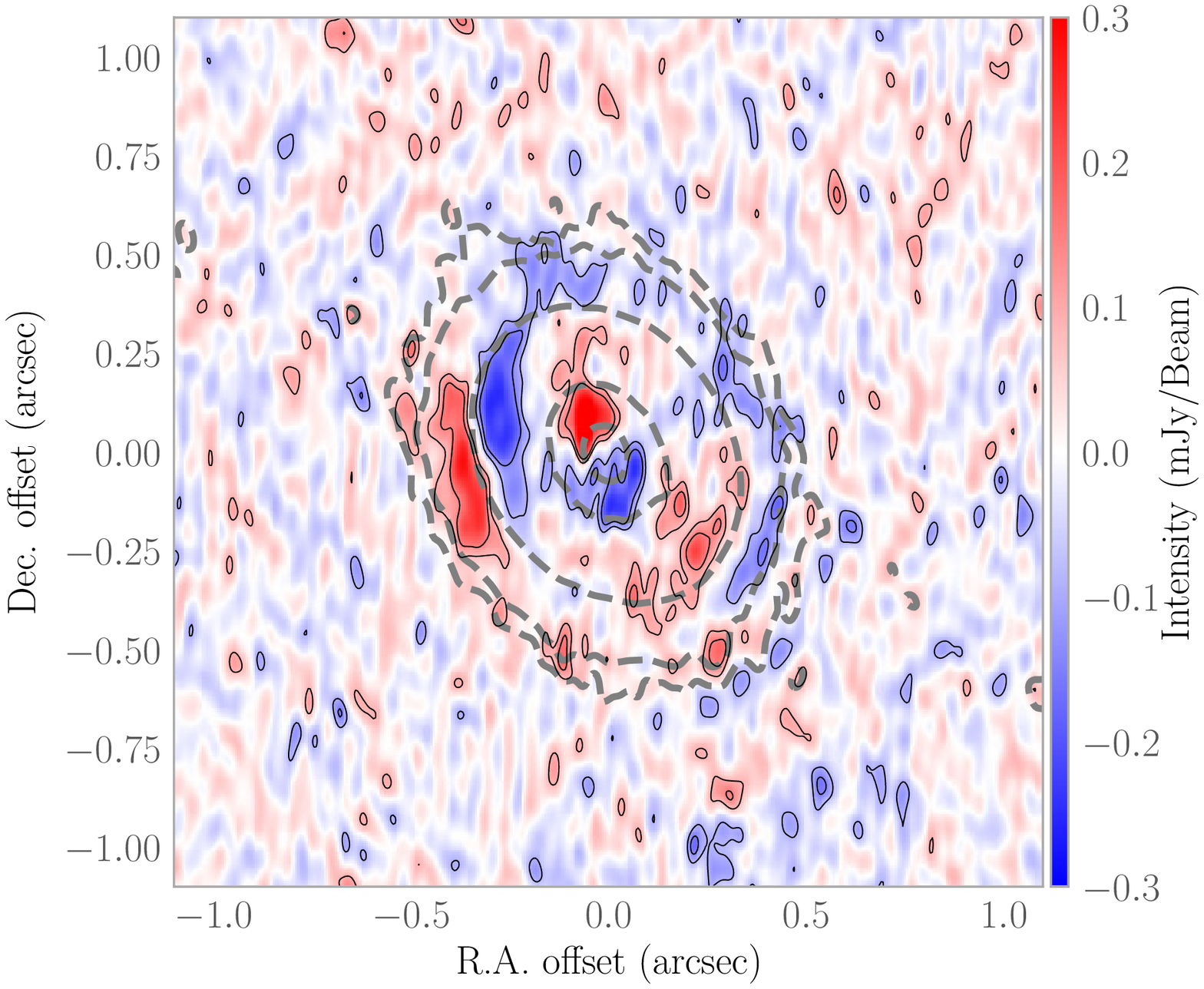}{0.49\textwidth}{(a) Residual at $233.0$~GHz}
\fig{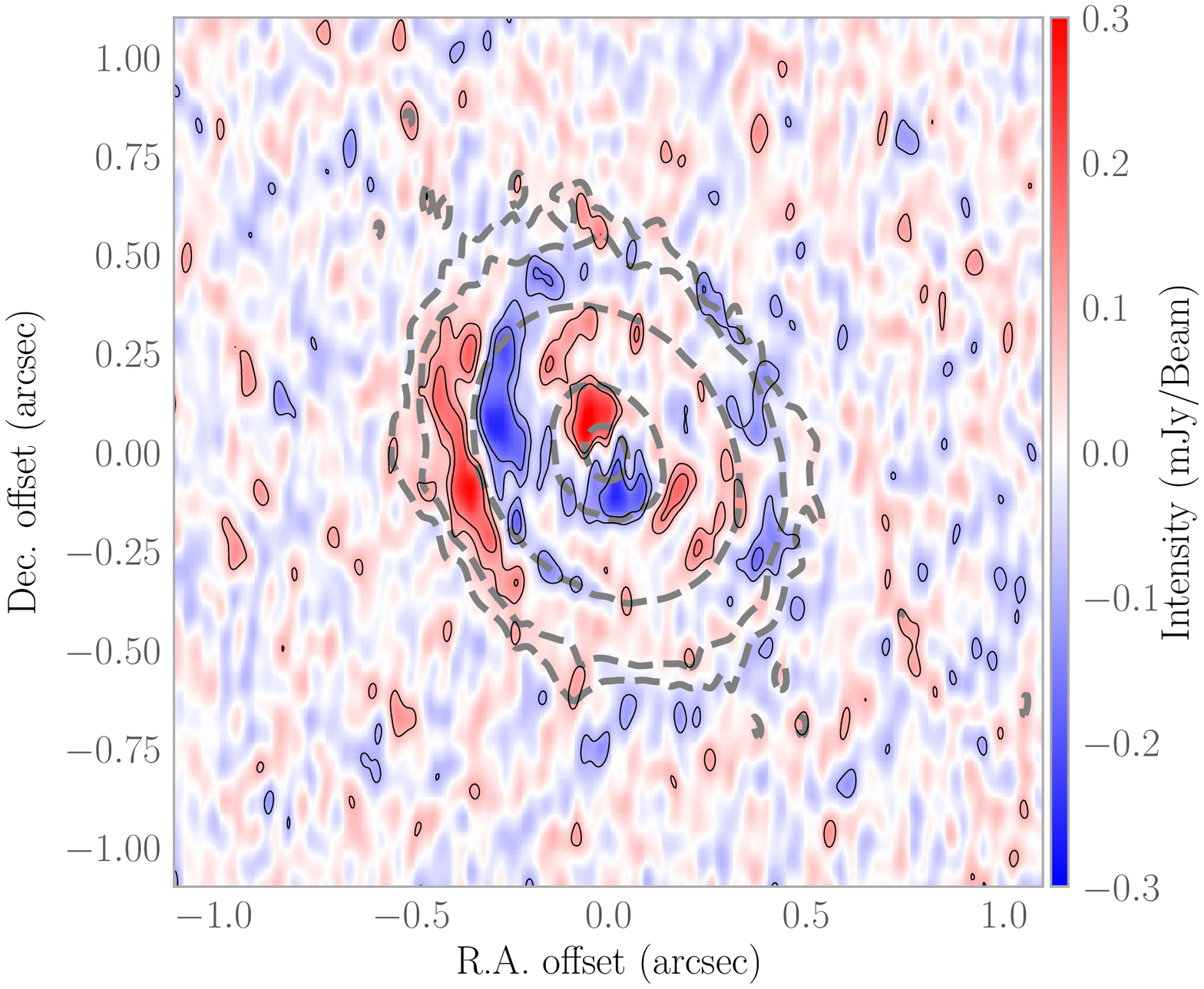}{0.49\textwidth}{(b) Residual at $216.7$~GHz}
}
\caption{
Residual between the observational data and model at 233.0~GHz (a), and at 216.7~GHz (b).
The contour indicates the levels of $\pm 3\sigma$ and $\pm 5 \sigma$.
The thick dashed contour lines indicate the observed intensity distribution, which are $0.087$~mJy/Beam ($3\sigma$), $0.29$~mJy/Beam ($10\sigma$), $2.9$~mJy/Beam ($100\sigma$), $8.7$~mJy/Beam ($300\sigma$), and $14.5$~mJy/Beam ($500 \sigma$) from the outside.
\label{fig:residual}}
\end{figure*}
Figure~\ref{fig:residual} shows the map of the residual between the model and the observation in the upper and lower band data.
The pattern of the residual in the upper and lower band data are similar to each other.
One can see significant residuals around the center, which is also shown by Figure~\ref{fig:comp_flux}.
Moreover, the residual map shows positive and negative structures at the upper left (R.A. offset $\simeq -0.4$~arcsec, Dec. offset $\simeq 0.2$~arcsec).
The amplitudes of those structures are larger than $5\sigma$, which can indicate the real asymmetric structures.

\subsection{Asymmetric structure} \label{subsec:asymmetricity}
As shown in the previous subsection, the residual map between the model and the observation indicates the asymmetric structures at the center and the outer disk.
Here we further investigate this asymmetry of the disk, by using a model-independent analysis.

In the visibility domain, the visibility is expressed by
\begin{align}
V(\rhovec) &= \int \int I(\rvec) e^{-j \rvec \cdot \rhovec} d\rvec,
\label{eq:visibility}
\end{align}
where $j=\sqrt{-1}$ is the imaginary unit and $\rhovec$ and $\rvec$ indicate the position vectors in the deprojected uv plane and the image, $I(\rvec)$ is the intensity distribution.
When $I(\rvec)$ is axisymmetric, we can express the visibility as
\begin{align}
V(\rho) &= 2\pi \int I(R) J_0(R\rho) R dR,
\label{eq:visibilty_symmetric}
\end{align}
where $J_0(k)$ is 0th-order Bessel function of the first kind.
The visibility of the axisymmetric disk has only a real part.
The image does not change if the disk is $180^{\circ}$ rotated against the disk center.
In mathematics, a $180^{\circ}$ rotated image has the visibility which is the complex conjugate of that of the original image.
Hence, the difference between the original and $180^{\circ}$ rotated images has only imaginary part, namely, twice the imaginary part of the original image.
When the system does not have a significant asymmetric structure, the difference is almost zero because the imaginary part of the visibility of the original image is very small.
On the other hand, when the disk has asymmetric structures, we could see some residual between the original and $180^{\circ}$ rotated images, which corresponds to the imaginary part of the visibility.
This approach of investigating asymmetric structures is totally model-independent.

\begin{figure}
\plotone{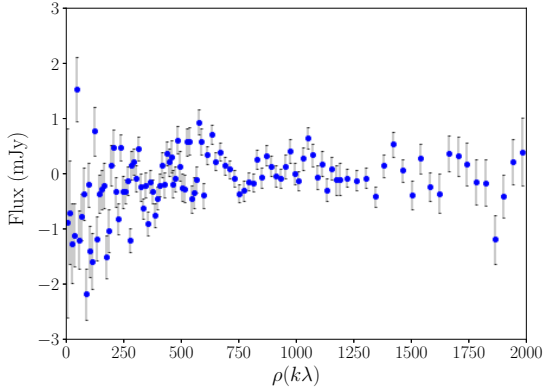}
\caption{
Imaginary part of the visibility combined from all data.
\label{fig:vis_imag}}
\end{figure}
\begin{figure}
\plotone{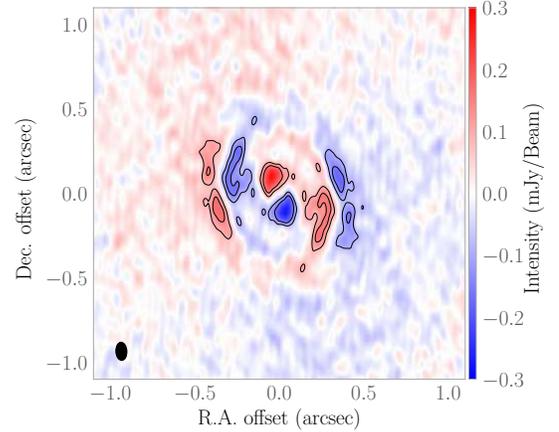}
\caption{
Image synthesized from the imaginary part of the visibility shown in Figure~\ref{fig:vis_imag}.
The contour indicates the levels of $\pm 3\sigma$ ($\pm 0.087$~mJy/Beam) and $\pm 5 \sigma$ ($\pm 0.145$~mJy/Beam).
\label{fig:diff}}
\end{figure}
We produced the synthesized image using all the data, namely, the C43-4 and C43-8 data in upper and lower bands.
Figure~\ref{fig:vis_imag} shows the azimuthally averaged imaginary part of the visibility.
The average is calculated in the same way as in Figure~\ref{fig:vis_diffband}.
The imaginary part of the visibility is much fainter than the real part.
However, one can see the significant signals as large as a few mJy, at $\rho \lesssim 300k\lambda$, which implies the asymmetric structure with the scale of $\gtrsim 0.7$~arcsec ($\gtrsim 140$~au).
Figure~\ref{fig:diff} shows the synthesized image produced only from the imaginary part of visibility data.
We can find the asymmetric structure at the center and that at the upper left (R.A offset $\simeq -0.4$~arcsec ($\simeq -80$~au) and Dec. offset $\simeq 0.2$~arcsec ($\simeq 40$~au)), which is consistent with the difference btween the observed image and the axisymmetric model shown in Figure~\ref{fig:residual}.
In addition, we see structures at the lower right, namely, R.A offset $\simeq 0.4$~arcsec and Dec. offset $\simeq -0.2$~arcsec.
The Fourier transform of pure imaginary function must be an odd function, which means that there is a counter part in the opposite location in the synthesized image.
If the asymmetric structure indicated by the imaginary part is real, we should find a signal at the same location with the model-subtracted image.
Hence, we consider that the structure at the lower right could be the counter part of the structure at the upper left.

As described in Section~\ref{sec:obs} and in the result shown above, we determined the phase center by the Gaussian fitting with CASA tool \verb#uvmodelfit#.
However, the above analysis of the imaginary part of the visibility depends on the choice of the phase center.
\begin{figure*}
\resizebox{0.98\textwidth}{!}{\includegraphics{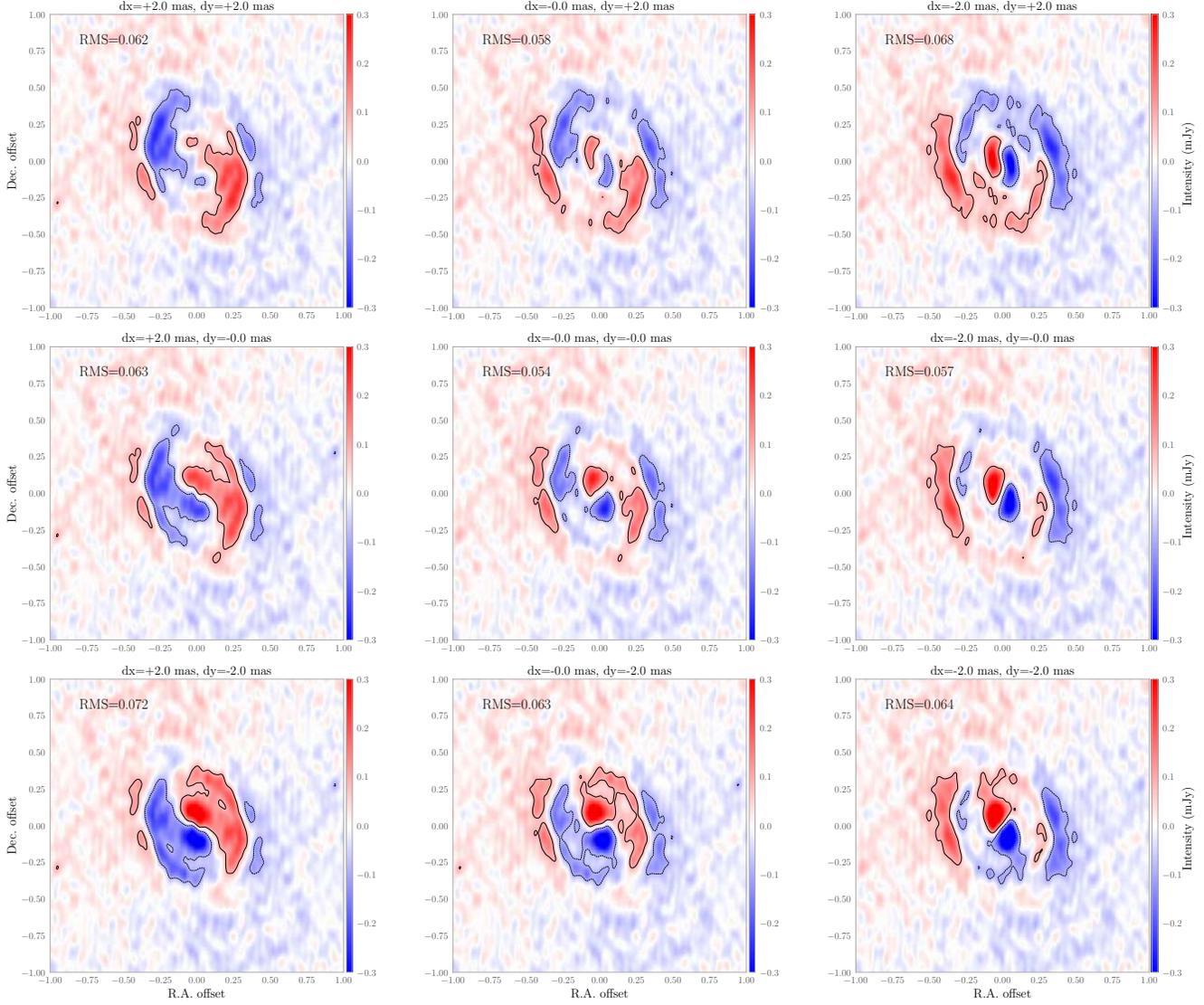}}
\caption{
Images deprojected only from the imaginary part of the visibility as shown in Figure~\ref{fig:diff}, but with the different phase center.
The central figure is the same as that shown in Figure~\ref{fig:diff} with the same phase center.
The horizontal coordinate of phase center is shifted in 2~mas ($-2$~mas) from that of the central figure, in left (right) column, and  the vertical coordinate of the phase center is shifted in 2~mas ($-2$~mas) in the top (bottom) row.
In the middle column (row), the horizontal (vertical) coordinate of the phase center is the same as that of the center figure.
\label{fig:imag_map}
}
\end{figure*}
Figure~\ref{fig:imag_map} shows how the images (deprojected to the disk plane) produced from pure imaginary visibility change with the choice of the phase center.
In the figure, we shift the center in $\pm 2$~mas ($\pm 0.38$~au) in horizontal and vertical directions from the phase center of  Figure~\ref{fig:diff}.
The image is shifted in the visibility domain by the phase shift defined as $\exp\left[2\pi\left(u dx+v dy\right)\right]$, where $u$ and $v$ are the spatial frequencies and $dx$ and $dy$ are shift values in R.A and Dec. directions, respectively. 
It is reasonable to assume that the disk structure is almost axisymmetric, despite the disk has some asymmetric structures.
Under this assumption, the image deprojected from the imaginary part with the 'correct' phase center has the minimum root mean square of the intensity.
We calculated the sum of the root mean square value of the intensity at each pixel within the radius of 0.6~arcsec (114~au) from the center, which is labeled at the upper left corner at each panel (labeled by RMS).
The figure with our fiducial phase center has the minimum value of RMS among the listed panels.
Hence, the phase center determined by the Gaussian fit is consistent with the center which minimalize the asymmetry.
On the other hand, in the panel at the upper left corner (dx=2~mas and dy=2~mas), the asymmetric structure around the center is almost vanished.
This indicates that the center of the inner structure is shifted to the center of the outer structure.

\section{Modeling by radiative transfer simulations} \label{sec:rt}
\subsection{Setup and model description}
We now have a model for the intensity distribution of the disk around WW~Cha.
In this section, we discuss the physical condition (e.g., dust surface density, size distribution of the dust grains, and temperature) of the disk, by using radiative transfer simulations with RADMC-3D \footnote{\url{http://www.ita.uniheidelberg.de/~dullemond/software/radmc-3d/index.html}} \citep{RADMC3D}.
We setup a model of an axisymmetric dust surface density distribution which is motivated by the intensity distribution derived in the previous section as
\begin{align}
&\rhosurf(R) =\Sigma_0 \left[ f_{I}\bracketfunc{R}{\rchara}{-s} \exp\left[ -\bracketfunc{R}{\rchara}{\zeta}\right]  \right. \nonumber \\
& \qquad \qquad + \left. \sum_{i=1}^{2} \rhi{i} \exp \left[-\bracketfunc{R - \rci{i}}{\rwi{i}}{2}\right] \right],
\label{eq:rhosurf}
\end{align}
Here $f_I$ is defined by Equation~(\ref{eq:f}), $\zeta$,$R_{\rm c}$, $H_i$,$W_i$, $R_{\rm rc,i}$, and parameters in $f_I$ ($\delta_{\rm cav}$,$\delta_{\rm gc,1}$,$\delta_{\rm gc,2}$, $R_{\rm cav}$, $R_{\rm g,in,1}$,$R_{\rm g,out,1}$) are fixed to the values given in Table~\ref{tab:bestfit} (for C43-8 data).
For the parameter $s$ that is related to the slope of the dust surface density, we use $s=0$, which is different from the best-fit parameter of $\gamma$ which is $\sim 0.35$.
We confirm that the choice of $s$ hardly affects the estimates of physical parameters.
When we use $s=0.5$, the mid-plane temperature is affected only by a few Kelvin.
The parameter $\Sigma_0$ controls the total mass of the dust $M_{\rm dust}$ and when $\Sigma_0=1.5\mbox{ g/cm}^2$, $M_{\rm dust}=3\times 10^{-3} M_{\odot}$.

We first vary the stellar luminosity and $\Sigma_0$ and check the agreement with observations in order to address the uncertainty of the estimate of the disk physical parameters.
We then vary the dust size in order to address the spectral index distribution.
Here we present a physical disk and star model that reasonably matches observations.
Full modeling studies that derive the uncertainties of all the parameters are beyond the scope of this paper.

Rest of the setup of the simulation is as follows.
In the vertical direction, we adopt a Gaussian shape distribution, namely,
\begin{align}
\rho(R,z) &= \frac{\rhosurf (R)}{\sqrt{2\pi}\hscale} \exp\left( -\frac{z^2}{2\hscale^2} \right),
\label{eq:rhodens}
\end{align}
where $\hscale$ is the scale height of the dust layer.
The dust scale height can be smaller than that of gas structure, due to dust settling \citep[e.g.,][]{Nakagawa_Sekiya_Hayashi1986}.
Hence, we consider two dust components:one is a small dust component with size distribution $\propto s^{-3.5}$, where $s$ is the size of the grains, and the minimum and maximum sizes are $0.1 \ \mu \mbox{m}$ and $0.1 \ \mbox{mm}$, respectively.
The other is a large dust component with its size having Gaussian distribution in logarithmic space.
The peak of the size distribution $\slarge$ is $1$~mm with the smallest size $0.6$~mm and with the largest size of $1.7$~mm.
We assume that the scale height of the large grains is 0.1 times the scale height of the gas.
The mass ratio between the small and large grains is set to be 1:9.
We adopt the same compositions of the dust grains to that adopted in \cite{DSHARP_DUST}\footnote{The optical constant file was provided by Dr. Ryo Tazaki.}.
The absorption and scattering coefficients for each component are averaged by the size distribution.

The mass and the surface temperature of the central star are set to $1.2\msun$ and 4350~K, respectively \citep{Luhman2007}.
By the SED modeling, the stellar luminosity is estimated by $11L_{\odot}$ \citep{Garufi2020}.
Since WW~Cha is a young newborn star, however, it could be still embedded in the core \citep{Ribas2013,Garufi2020}.
In such a case, the luminosity may be underestimated due to high extinction \citep{Follette_OphIRS48_2015}.
Hence, we also consider the case with $L_{\ast}=22L_{\odot}$.

The radial coordinate extends from $0.1$~au to $1000$~au, which is divided into 256 meshes with logarithmic spacing.
The azimuthal angle $\theta$ and polar angle $\phi$ are divided into 256 meshes in $0<\theta < 2\pi$ and in $0<\phi<\pi$ (the midplane is located at $\phi=\pi/2$), respectively.
We adopted $3\times 10^{8}$ photons for thermal Monte Carlo radiative transfer and imaging, and for the SED, $10^{7}$ photons are adopted.

We first carried out radiative transfer simulations with the disk scale height calculated by the empirical formula given by \cite{Dong_Najita_Brittain2018}, namely, $T=220(L_{\ast}/11L_{\odot})^{1/4}(R/\rm{1au})^{-0.5}$ (it is quite similar to Equation~\ref{eq:tmid}).
After the first run, we calculate the temperature on the midplane at each radius and we preformed simulation again with the scale height given by the midplane temperature.
We repeated the above cycles until the temperature distribution is converged.
In our case, the temperature distribution is converged after 2--3 iterations.

For the imaging, we converted the output of RADMC-3D to the ALMA measurement set with observed measurement set by use of \verb#vis_sample#.
Then, we deprojected the image from the model measurement set with the imaging parameter listed in Table~\ref{tab:obs}.

\subsection{Results} \label{subsec:rt_result}
\subsubsection{Intensity and spectral energy distribution}
\begin{figure}
\plotone{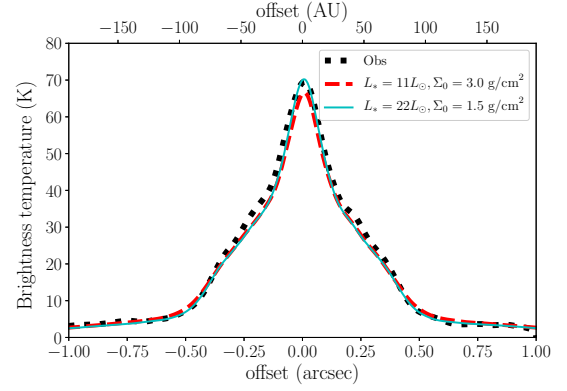}
\caption{
Brightness temperature distributions along the major axis of the disk, for the observation (black dot), the results of the simulations with $L_{\ast}=11L_{\odot}, \Sigma_0=3\mbox{ g/cm}^2$ (red dashed) and with $L_{\ast}=22L_{\odot}, \Sigma_0=1.5 \mbox{ g/cm}^2$ (cyan solid).
\label{fig:comp_tb_mcrt}}
\end{figure}
Figure~\ref{fig:comp_tb_mcrt} compares the brightness temperatures given by the observations and simulations.
With $L_{\ast}=11L_{\odot}$, we need $\Sigma_0=3\mbox{ g/cm}^2$ which corresponds to the total dust mass of $7\times10^{-3} M_{\odot}$. 
In this case, the disk is highly gravitationally unstable in most part if the gas-to-dust mass ratio is 100, as shown below.
If the stellar luminosity is larger by a factor of two ($L_{\ast}=22L_{\odot}$), we found that $\Sigma_0=1.5 \mbox{ g/cm}^2$ is enough to reproduce the observation.

\begin{figure}
\plotone{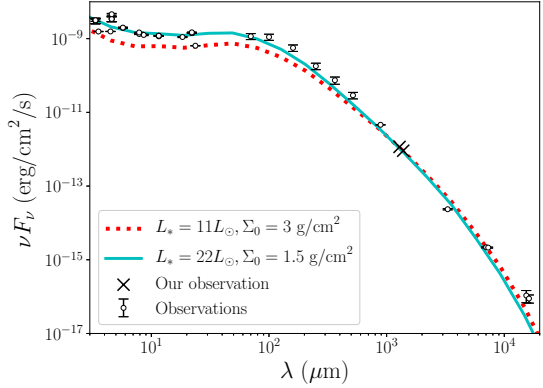}
\caption{
Spectral energy distribution given by observations (dots) and the simulations with $L_{\ast}=11L_{\odot}, \Sigma_0=3\mbox{ g/cm}^2$ (red dashed) and with $L_{\ast}=22L_{\odot}, \Sigma_0=1.5 \mbox{ g/cm}^2$ (cyan solid).
The crosses indicate the total fluxes given by our observation at 233~GHz (494 mJy) and 216.7~GHz (418 mJy).
The observational data with $\lambda < 1$~mm are extracted from VizieR database (\url{https://vizier.u-strasbg.fr/}) and the references are in the main text.
\label{fig:sed}}
\end{figure}
Figure~\ref{fig:sed} illustrates the SED at $0.3\mu \mbox{m}$ -- 2~cm, given by the previous observations \citep{Lommen2007,Lommen2009,Gutermuth2009,Ishihara2010,Cutri2014,Pascucci2016,Ribas2017}, including our result, and the simulations.
Both simulations with $L_{\ast}=11L_{\odot}$ and $22L_{\odot}$ can reproduce the ALMA band~6 flux ($\sim 230$~GHz).
For fluxes at the longer wavelengths, namely $3$~mm $<\lambda < 1$~cm, simulations agree with the observation.
The fluxes of the simulations at $\lambda > 1$~cm is smaller than the observed flux, though it could be due to the contribution from the free-free emission from the star \citep{Rodmann2006}.
In the case with $L_{\ast}=11L_{\odot}$, the flux at far infrared wavelengths is about a factor of two smaller than the observed values.
In this case, we need some contribution from the envelope to account for infrared flux.
On the other hand, the case with $L_{\ast}=22L_{\odot}$ can reproduce the fluxes from the infrared to the radio, by only the emission from the star.

\subsubsection{Dust density and temperature}
\begin{figure}
\plotone{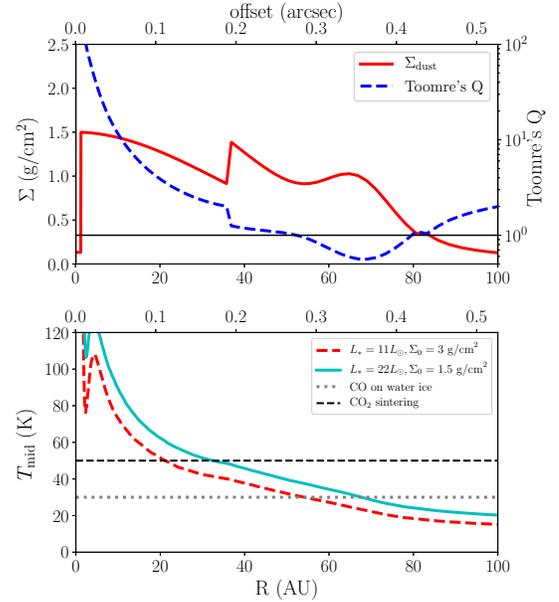}
\caption{
({\it Upper})
Dust surface density and Toormre's Q-value (with gas-to-dust ratio being 100) in the case of $\Sigma_0=1.5\mbox{ g/cm}^2$ (in the case of $\Sigma_0=3 \mbox{ g/cm}^2$, $\Sigma$ simply becomes two times larger and Q-value decreases with the increase of $\Sigma$.).
The horizontal thin line indicates $Q=1$.
({\it Lower})
Midplane temperatures given by the radiative transfer simulations.
The two horizontal lines indicate $50$~K and $30$~K from the top.
\label{fig:dens_midtemp}}
\end{figure}
The stellar luminosity affects estimate on the mass of the gas disk, but it does not significantly affect the midplane temperature.
In the upper panel of Figure~\ref{fig:dens_midtemp}, when $\Sigma_0=1.5 \mbox{ g/cm}^2$, we show the distribution of the dust surface density and Toomre's Q-value \citep{Toomre1964}, assuming the gas-to-dust ratio of 100.
The lower panel of Figure~\ref{fig:dens_midtemp} shows the midplane temperatures given by simulations.
The midplane temperature with $L_{\ast}=22L_{\odot}$ is just about 1.16 times higher than that with $L_{\ast}=11L_{\odot}$, roughly corresponding to $L_{\ast}^{1/4}$ dependence as expected from Equation~(\ref{eq:tmid}).

We find that the disk is expected to have relatively low values of Toomre's Q-value.
As shown in the upper panel of Figure~\ref{fig:dens_midtemp}, in the case of $\Sigma_0=1.5\mbox{ g/cm}^2$, the Q-value is smaller than or close to unity at 20~au $<R<$ 100~au.
In particular, the dust bump ($40$~au -- $70$~au) is gravitationally unstable because $Q\lesssim 1$ and the dust bump may fragment.
In the case of $\Sigma_0=3 \mbox{ g/cm}^2$, the Q-value are decreased by a factor of two.
In this case, the disk is still nearly gravitationally unstable and the fragmentation is expected at $R>10$~au and turbulence due to gravitational instability is also expected.
However, no fragment-like structure is seen in the dust-continuum image.
This may indicate that the gas-to-dust ratio is smaller than 100 in the region with $>20$~au.
It is discussed in Section~\ref{subsec:origin_substr}.

We note that the temperatures at the peak locations are $\sim 30$ -- $50$~K, which is close to the freezeout temperature of the CO on water ice.
This may be related to the origin of the bump structure, as discussed in Section~\ref{subsec:origin_substr}.

\subsubsection{Spectral index} \label{subsec:spectral_index}
\begin{figure*}
\gridline{\fig{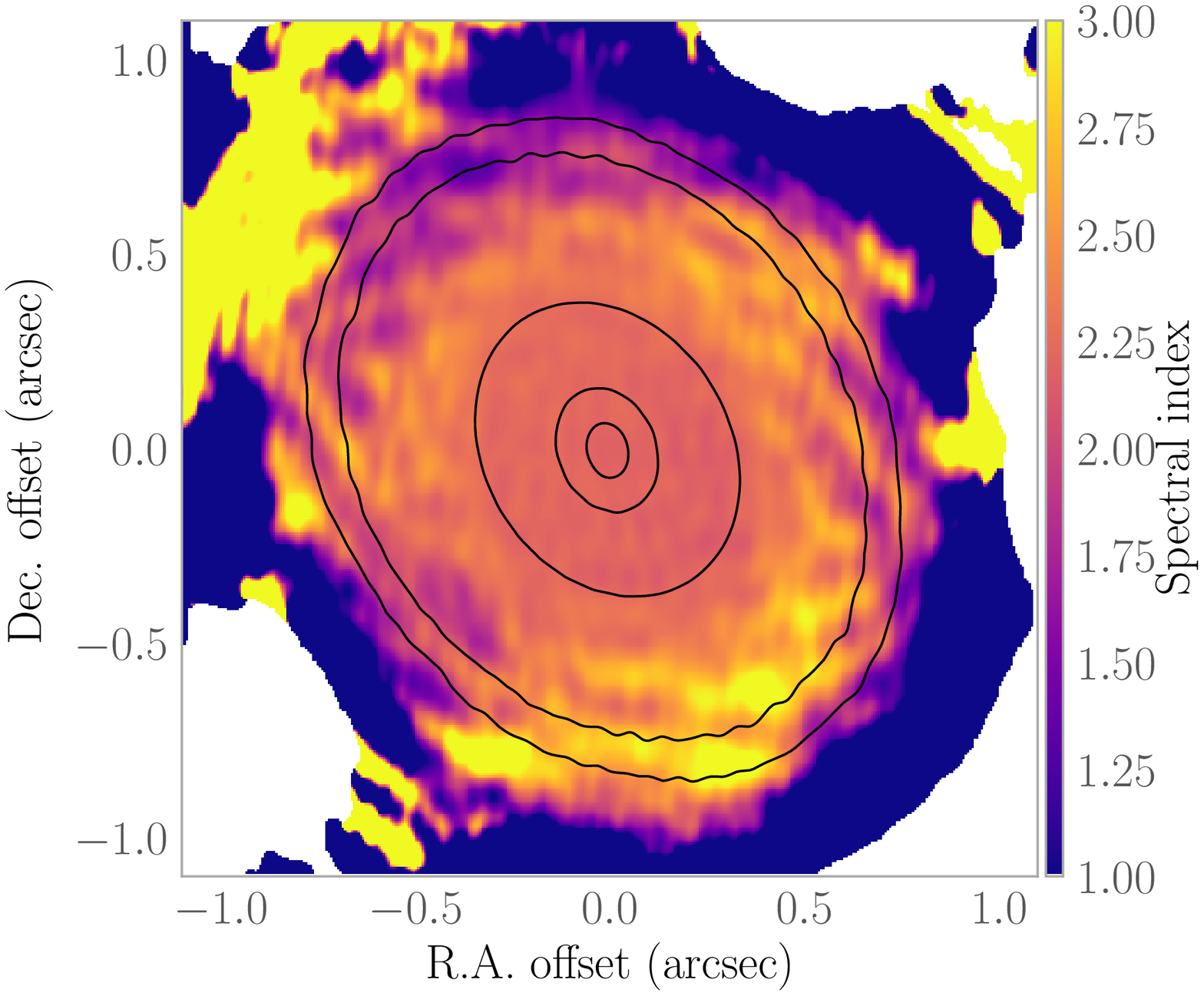}{0.49\textwidth}{(a) $s_{\rm large}=1.0 \mbox{ mm}$}
\fig{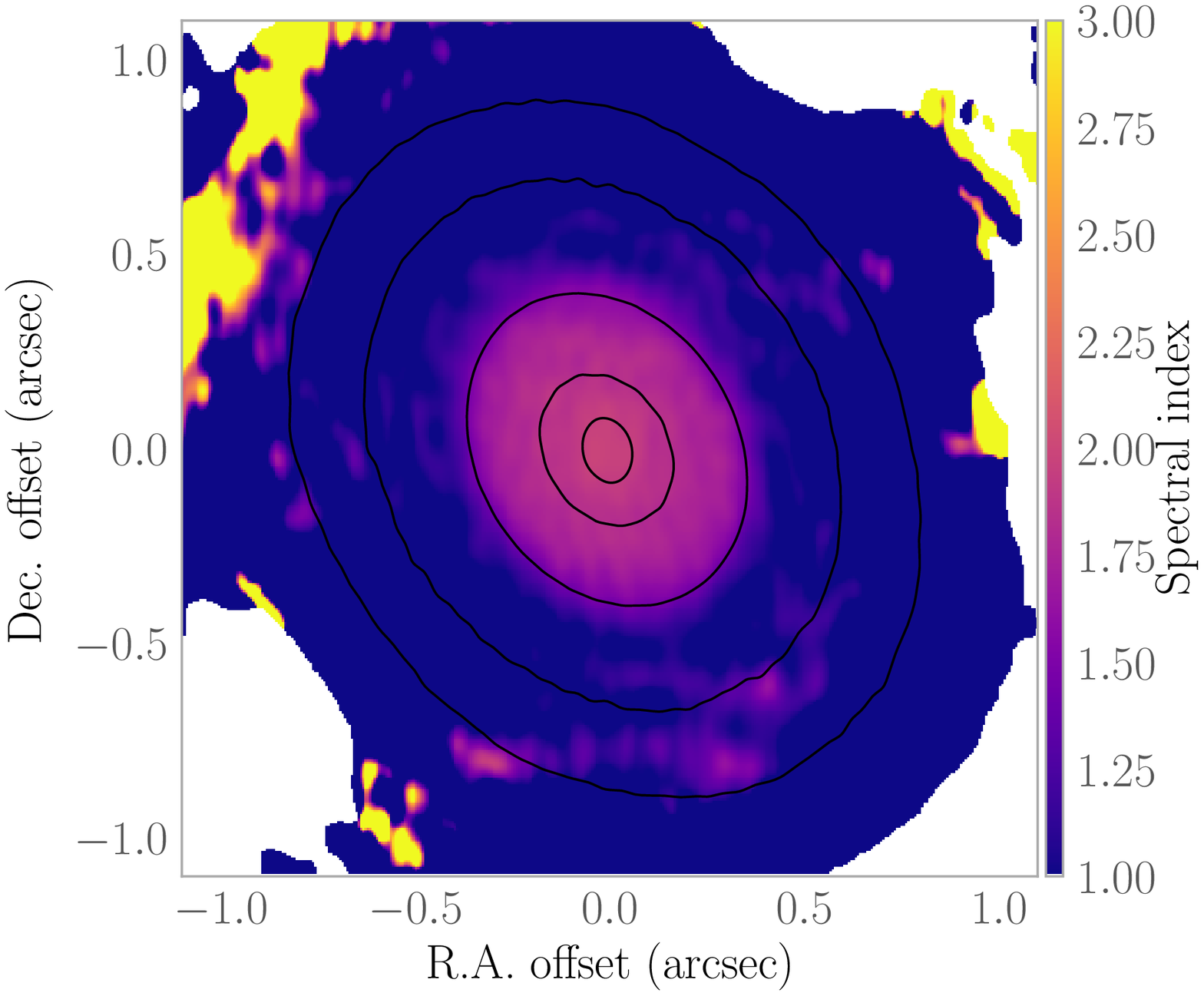}{0.49\textwidth}{(b) $s_{\rm large}=0.5 \mbox{ mm}$}
}
\caption{
Spectral index maps given by radiative transfer simulations, with $\slarge=1\mbox{ mm}$ (left panel) and $\slarge=0.5\mbox{ mm}$ (right panel).
The contours indicates the same flux levels shown in the panel (b) of Figure~\ref{fig:obs}.
\label{fig:spectral_index_mcrt}}
\end{figure*}
Using the images given by the simulations at upper and lower bands, we produced the map of the spectral index by the use of \verb#tclean# task with \verb#nterm#=2.
Although it is calculated from the very narrow frequency range, we may constrain the dust size distribution from it.
The spectral index depends on the peak size of the large grain $\slarge$, rather than stellar luminosity and the dust surface density.
Hence, we carried out additional simulations with $\slarge = 0.5 \mbox{ mm}$.
In Figure~\ref{fig:spectral_index_mcrt}, we show the spectral index map in the case of $\slarge=1.0 \mbox{ mm}$ and $0.5 \mbox{ mm}$, when $L_{\ast}=22L_{\odot}$ and $\Sigma_0=1.5 \mbox{ g/cm}^2$.
In the case of $\slarge=1.0 \mbox{ mm}$, the spectral index increases in the outer region, which agrees with the observations (panel (b) of Figure~\ref{fig:obs}), but it is larger than 2 around the center.
In the case of $\slarge=0.5 \mbox{ mm}$, on the other hand, the spectral index is below 2 around the center, whereas it is much smaller than the observed index at the outer region.

\begin{figure}
\plotone{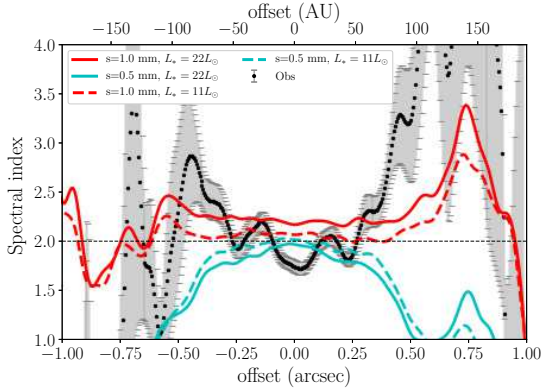}
\caption{
Distributions of the spectral index along the major axis in the case of $\slarge=1.0\mbox{ mm}$ (red) and $0.5\mbox{ mm}$ (cyan).
In the upper panel, $L_{\ast}=22L_{\odot}$ and $\Sigma_0=1.5\mbox{ g/cm}^2$, while in the lower panel, $L_{\ast}=11L_{\odot}$ and $\Sigma_0=3.0\mbox{ g/cm}^2$.
The dots indicates the observed one shown in the panel (d) of Figure~\ref{fig:obs}.
\label{fig:spectral_index_slice_mcrt}}
\end{figure}
Figure~\ref{fig:spectral_index_slice_mcrt} compares the distributions of the spectral index along the major axis between the simulations and the observation.
As can be seen in the figure, the distributions are similar between the case with $L_{\ast}=22L_{\odot}$ and $\Sigma_0=1.5\mbox{ g/cm}^2$ and the case with $L_{\ast}=11L_{\odot}$ and $\Sigma_0=3.0\mbox{ g/cm}^2$.
In both the cases, for the outer region of offset $>0.25$~arcsec, the observational feature that the spectral index increases in the outer region is consistent with the radiative transfer calculations when $\slarge=1.0 \mbox{ mm}$.
For the inner region of $<0.25$~arcsec, the spectral index below 2 is consistent with the case of $\slarge=0.5 \mbox{ mm}$.

The spectral index also depends on the size distribution of the large grains, though the intensity does not significantly depends on that.
In Appendix~\ref{sec:rt_powd}, we demonstrate radiative transfer simulations with the large grains which have a power-law size distribution like that of the small grains.
With the power-law distribution, the spectral index does not become below 2, even if the maximum size of the dust grains is 0.5~mm.
Hence, the log-normal size distribution of large grains may be preferred for the inner region.
On the other hand, the power-law distribution may be preferred for the outer region as can be seen in Figure~\ref{fig:alpha_slice_mcrt_powd}.

We should note that the spectral index discussed in this paper is provided from the very narrow range of the observed frequency within band~6.
The spatial variation of spectral index should be investigated by future observations at multiple wavelengths.
In Appendix~\ref{sec:alpha_diff_bands}, we show a few examples of the spectral index based on ALMA bands, which are calculated from the results of radiative transfer simulations.
The spectral index can be different by the choice of the bands which are taken to calculate the spectra index.
We may be able to constrain the dust size distribution from the spectral indexes in multiple bands.

\section{Discussion} \label{sec:discussion}
\subsection{Origin of substructures} \label{subsec:origin_substr}
\subsubsection{Ring} \label{sss:ring}
We found the bump with double peaks at 40 -- 80 au from the central star by the model fitting in the visibility domain.
One of the peaks is at $\sim 40$~au and the other is at $\sim 70$~au.

The bump structure can be formed by dust trapping due to the pressure bump \citep[e.g.,][]{Pinilla_Birnstiel_Ricci_Dullemond_Uribe_Testi_Natta2012, Dullemond_DSHARP}.
However, the bump that is formed by the above mechanisms likely has a single peak, while the bump of our model has double peaks.
If there are two pressure bumps, it might explain a shape with double peaks.
Another possible location where a dust bump is likely to form is the outer edge of the planet-induced gap \citep[e.g.,][]{paardekooper2004,Muto_Inutsuka2009b,Zhu2012,Pinilla_Ovelar_Ataiee_Benisty_Birnstiel_Dishoeck_Min2015,Dong_Zhu_Whitney2015,Kanagawa_Muto_Okuzumi_Taki_Shibaike2018}.
However, we did not detect any gap structure interior to the bump structure.
Hence, it might not be the structure induced by the dust trap of the pressure bump and planetary gap.

One plausible scenario of producing a bump with double peaks uses the snowline.
The bump can be formed due to volatile freeze-out altering the coagulation and fragmentation of dust grains \citep[e.g.,][]{Zhang_Blake_Bergin2015,Okuzumi_Momose_Sirono_Kobayashi_Tanaka2016}.
As shown in Figure~\ref{fig:dens_midtemp}, the temperature around the outer peak is about $30$~K, which is close to the freezeout temperature of CO at water ice \cite{Huang_DSHARP}.
Around the inner peak, the temperature is about $50$~K.
Although it is not close to the condensation temperatures of major volatiles (e.g., CO, CO$_2$), it is close to the condensation temperature of H$_2$S \citep{Zhang_Blake_Bergin2015}.
Moreover, in $T\sim 50$~K, the sintering can occur for some species such as H$_2$S and C$_2$H$_8$ \citep{Okuzumi_Momose_Sirono_Kobayashi_Tanaka2016}.
Hence, this bump with double peaks may be formed by the snowline and the sintering effect.

\subsubsection{Asymmetric structure} \label{sss:asymmetric_str}
As discussed in Section~\ref{subsec:asymmetricity}, the asymmetric structure is suggested at the outer region of the disk, the positive (bright) structure at R.A. offset $\simeq -0.5$~arcsec and the negative (dark) structure at R.A. offset $\simeq -0.3$~arcsec. 
\begin{figure}
\plotone{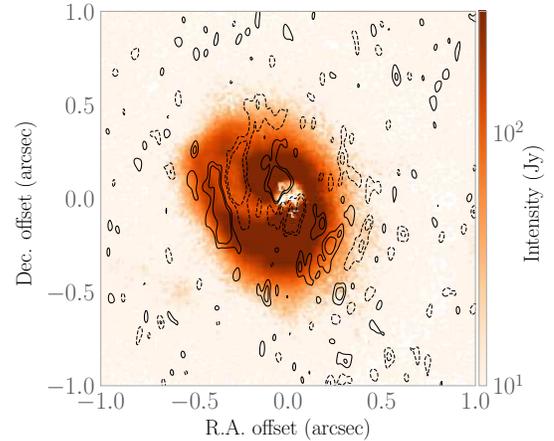}
\caption{
\label{fig:vs_sphere}
Comparison between the SPHERE image \citep{Garufi2020} (color) and the residual map for the upper band data shown in Figure~\ref{fig:residual} (contour).
The contours show $\pm 3\sigma$ and $\pm 5\sigma$ levels.
The solid and dashed contours indicate positive and negative excesses, respectively. 
}
\end{figure}
Recently \cite{Garufi2020} have observed a bright spiral in the disk of WW Cha by SPHERE observation, which wraps around from east to north in clockwise.
In Figure~\ref{fig:vs_sphere}, we compare the infrared image given by SPHERE and the image synthesized from the imaginary part of the visibility (the same as that shown in Figure~\ref{fig:diff}).
The spiral feature in the SPHERE image has bright and faint parts.
The location of the positive asymmetric feature in sub-mm observations coincides with the bright NIR spiral while that of the negative asymmetric feature coincides with the location of faint spiral region of the NIR observation.

In tandem with the radiative transfer modeling, we suggest that the spiral features are formed by gravitational instability.
When the value of Toomre's Q is smaller than $\sim 1$, the effects of self-gravity become prominent spiral features appear \citep[e.g.,][]{Lodato_Rice2004}.
As shown in Figure~\ref{fig:dens_midtemp}, hence, the disk can be gravitational unstable at $R=40$ -- $80$~au, when $\Sigma_0=1.5 \mbox{ g/cm}^2$, if the gas-to-dust ratio is 100.
When the disk was highly gravitationally unstable, we could observe clear spiral waves or fragmented structures.
However, we cannot see clear significant spirals or fragments on the disk, except the relatively weak structure at the upper left.
Hence, the gas-to-dust ratio may be smaller than 100, especially within the bump region.
The faint positive structure at the upper left might be explained by the gravitational instability if the disk is marginally stable with a smaller gas-to-dust ratio.
Alternatively, we cannot see significant spiral pattern because the disk is optically thick.
In this case, the spiral might be observed at longer wavelengths.

Another possible mechanism for making asymmetry is dust concentration in the vortex.
Several protoplanetary disks have been observed to have vortex structures, for instance, HD~142527 \citep{Fukagawa2013,Soon2019}, Sz~91 \cite{Tsukagoshi2014,Canovas2016}, Oph IRS~48 \citep{Marel2015}, MWC~758 \citep{Dong2018_MWC758}, and etc.
One possible mechanism to form such a vortex is trapping dust grains into vortex formed by Rossby wave instability \citep[e.g.,][]{Lovelace_Colgate_Nelson1999,Li_etal2000,Lin2014,Fu2014,Ono2016}.
However, as different from the above disks, the asymmetric structure found in the disk of WW Cha is so faint that it is not visible in the raw image (Figure~\ref{fig:obs}).
It is visible only in the image with subtraction (the right panel of Figure~\ref{fig:residual} and Figure \ref{fig:diff}). 
The region of $R>0.3$~arcsec can be optically thin and the intensity of the asymmetric structure ($\sim 5 \sigma$) is just 5~\% of that of the symmetric structure ($\sim 100 \sigma$ at $R\simeq 0.4$~arcsec, as can be seen in Figure~\ref{fig:obs}).
Such a faint structure might be difficult to be formed by the dust trap into the vortex.

The origin of the negative asymmetric structure could not be explained by the above mechanisms.
Further theoretical models are required to discuss the origin of the negative asymmetric structure.

The asymmetric structure may be originated from the collapsing cloud/envelope.
The large-scale structure observed by e.g., SPHERE \citep{Garufi2020} indicate presence of the envelope, but the system is not significantly embedded as the extinction is not so large ($A_V\sim 4.8$~mag \citep{Ribas2013}, compared to e.g., $24$ -- $36$~mag of HL~Tau system \citep{ALMA_HLTau2015}).
Since the star has a high accretion rate, the asymmetry may be associated to the accretion variability or a jet.
So far, we do not see a significant jet-like structures both in the SPHERE and our dust continuum images so it is difficult to address this further with current observations.
Further observations (e.g., H$\alpha$ observations to see accretion variability or jet structure) would be required.

\subsection{Inner cavity and binary} \label{subsec:binary}
Although a large cavity is ruled out by the observed image shown in Figure~\ref{fig:obs}, our model allows the existence of a small cavity with the radius of about $1$~au.
However, further observations are required to confirm the small cavity, because it is much smaller than the angular resolution with the baselines of $<2000k\lambda$.
Moreover, we should note that there is a relatively large uncertainties on the MCMC fitting of the cavity size ($R_{\rm cav}$) and depth ($\deltahole$) as seen in Table~\ref{tab:bestfit}.
It indicates that our fitting cannot rule out the solution with no cavity.

If the disk has the small cavity, it can be formed by binary interaction \citep[e.g.][]{Artymowicz_Lubow1994,Dunhill_Cuadra_Dougados2015,Miranda_Munoz_Lai2017,Thun_Kley_Picogina2017,Price2018}.
The binary separation $a_{\rm bin}$ may be estimated from the cavity radius $R_0$ as $R_0=2.5 a_{\rm bin}$ \citep{Artymowicz_Lubow1994}, and equivalently $a_{\rm bin} \simeq 0.4$~au.
This separation is roughly consistent with the binary motion observed by \cite{Anthonioz2015}.

As shown in Section~\ref{subsec:asymmetricity}, the center of the inner structure can be different in $\sim 2$~mas from the center of the outer structure.
When the eccentricity of the binary motion is relatively large, the shape of the cavity induced by the binary interaction is also eccentric \citep[e.g.][]{Thun_Kley_Picogina2017}.
On the other hand, the shape of the outer disk can keep a symmetric shape because of the small binary separation. 
Hence, the binary eccentricity may be relatively large, if the star is binary.

\subsection{Dust size distribution} \label{subsec:dust_growth}
Finally, we discuss the size distribution of the dust grains in the disk from the spectral index map, though it is calculated from the narrow range of the observed frequency.
As can be seen in the panels (b) and (d) of Figure~\ref{fig:obs}, the spectral index below 2 around the center can be induced by the optical thick scattering emission \citep{Zhu2019,Liu2019}, and the model with $\slarge=0.5 \mbox{mm}$ can reproduce such a spectral index.
In the outer region of $>0.25$~arcsec from the center ($R >50$~au), the spectral index increases in the outer region, which is consistent with the model with $\slarge=1.0 \mbox{ mm}$.
This feature implies that the size of the large dust is larger in the outer disk.
Here, we discuss how such a distribution can be realized.
When the Stokes number of the dust grains is much smaller than unity, the radial drift velocity can be written by \citep[e.g.,][]{Nakagawa_Sekiya_Hayashi1986}
\begin{align}
\vrdust \simeq -2\st \eta \vk,
\label{eq:dust_drift_vel}
\end{align}
where $\vk$ denotes the Keplerian rotation velocity, $\st$ is the Stokes number of the dust grains given by $\pi\rho_d s_d/(2\Sigma_{\rm gas})$ ($\rho_d$ is the internal density of the dust), and $\eta=d\ln P/d\ln R$ is $\sim 10^{-3}$ in conventional disk models.
Our model indicates $\Sigma_{\rm dust} \sim 1.0 \mbox{ g/cm}^2$ (Figure~\ref{fig:dens_midtemp}), and hence, the Stokes number of 1~mm-sized dust is about $0.005$ when the gas-to-dust ratio is 100 and $\rho_d = 3$.
The radial drift timescale of the grains, $\tau_{\rm drift}=-R/\vrdust$, can be estimated by
\begin{align}
\tau_{\rm drift} \simeq & 5.5 \mbox{ Myr} \bracketfunc{\Sigma_{\rm dust}}{1 \mbox{g/cm}^2}{} \bracketfunc{\epsilon}{100}{}\bracketfunc{\rho_d}{3\mbox{ g/cm}^3}{-1} \nonumber \\
&\times \bracketfunc{s_{\rm dust}}{1\mbox{ mm}}{-1} \bracketfunc{\eta}{10^{-3}}{-1} \bracketfunc{R}{50\mbox{ au}}{3/2},
\label{eq:tau_drift}
\end{align}
where $\epsilon$ is the gas-to-dust ratio and $s_{\rm dust}$ is the size of the dust grain, respectively.
WW~Cha is young and the age is about $\lesssim 1$~Myr, which is shorter than the drift timescale of $1$~mm-sized dust and longer than the growth timescale of the dust, $\tau_{\rm growth} \simeq
\epsilon/\omegak = 0.005 \mbox{ Myr} (50 \mbox{ au}/R)^{3/2}$ with $\epsilon=100$ \citep{Brauer2008}.
Hence, the 1~mm-sized dust grains observed in the outer region is consistent with the dust drift.
In the inner region, the drift timescale of the dust grains becomes shorter.
Considering the star is as young as $<1$~Myr, the drift timescale can be comparable with the stellar age at $R< 16$~au.
This may explain the reason why the size of the dust grains is smaller than in the outer region.

The size of the grains can be determined by turbulent fragmentation \citep{Birnstiel_Dullemond_Brauer2010}.
In this case, the $\alpha$-parameter relevant to that the fragmentation threshold size equals to $\sim 1$~mm is 
\begin{align}
\alpha &=  1.57 \times 10^{-2} \bracketfunc{\Sigma_{\rm dust}}{1 \mbox{ g/cm}^2}{}\bracketfunc{\epsilon}{100}{} \nonumber \\
& \times \bracketfunc{\rho_d}{1\mbox{ g/cm}^3}{-1} \bracketfunc{s_{\rm dust}}{1\mbox{ mm}}{-1} \bracketfunc{v_f}{10 \mbox{ m/s}}{2} \bracketfunc{T}{50\mbox{ K}}{-1},
\label{eq:alpha}
\end{align}
where $v_f$ is the fragmentation threshold velocity.
Hence, considering the mid-plane temperature shown in Figure~\ref{fig:dens_midtemp}, we can estimate $\alpha\simeq 10^{-2}$.
This relatively large $\alpha$ is consistent with the high accretion rate onto the star \citep{Manara2016}, whereas it is larger than a value of $\alpha$ estimated by the recent observations of the disks around the aged stars, namely, HD~163296 \citep{Flaherty2015}, and TW~Hya \citep{Teague2016TWHydra,Flaherty2018TWHydra}, namely, $\alpha \lesssim 10^{-3}$.
Such a high viscosity may be due to turbulence induced by gravitational instability \citep[e.g.,][]{Boley2006}.

As mentioned in Section~\ref{subsec:spectral_index}, we note that the above discussion is based on the spectral index provided only from the very narrow range within band~6.
The spectral index should be investigated by future multi-wavelength observations.

\section{Conclusion} \label{sec:conclusion}
We presented the dust continuum emission of the protoplanetary disk around WW~Cha, observed by ALMA band~6.
Our conclusions are summarized as follows:
\begin{enumerate}
  \item  The dust continuum image clearly shows no large cavity, and a faint dust bump (Figure~\ref{fig:obs}).
  We also found the asymmetric structure at the center (Figure~\ref{fig:obs_slice}). Moreover, since the visibility is clearly different in the upper ($233.0$~GHz) and lower bands ($216.7$~GHz) (Figure~\ref{fig:vis_diffband}), we can obtain the spectral index map. The spectral index around the center is below 2, and it becomes larger in the outer region.
  \item  We constructed a model to fit the observation in visibility domain by MCMC fitting (Section~\ref{sec:modeling}). 
  Our model has a bump extending from $\sim 40$~au from the central star to $\sim 80$~au, with two local peaks located at $\sim 40$~au and $\sim 70$|au.
  As a result of radiative transfer simulations (Figure~\ref{fig:dens_midtemp}), the midplane temperature around the outer peak is close to the freezeout temperature of CO on water ice ($\sim 30$~K).
   The midplane temperature around the inner peak is about $50$~K, which is close to the condensation temperature of H$_2$S and it is also close to the temperature that the sintering can be caused for several species. 
   Hence, this bump may be induced by the snowline and the sintering effect.
  \item The residual map between the observed data and best fit model indicates asymmetric structures at the center and the upper left of the disk. 
  We also confirmed those asymmetric structures by a model-independent method, which is imaging of the imaginary part of the visibility of the observed data (Section~\ref{subsec:asymmetricity}). 
  These structure could be robust, though the amplitude of the asymmetric structures are faint ($\sim 5\sigma$ level ) as compared with the symmetric structure.
  \item The spectral index map given by the observation may be consistent with the result of radiative transfer simulations with the relatively large dust grains (1~mm) in the outer region, whereas the result of the simulation with the smaller dust grains (0.5~mm) can be suitable for the region close to the center (Figures~\ref{fig:spectral_index_mcrt} and \ref{fig:spectral_index_slice_mcrt}). This implies that the size of the dust grains is larger than that in the inner region. As discussed in Section~\ref{subsec:dust_growth}, such a distribution is consistent with the radial drift and collisional growth of the dust grains, because of the massive disk around a young star.
\end{enumerate}

We thank Ryo Tazaki for providing the optical constant data to calculate opacity of dust grains.
KDK was also supported by JSPS Core-to-Core Program ``International Network of Planetary Sciences'' and JSPS KAKENHI. 
This work is in part supported by JSPS KAKENHI grant Nos. 18H05441 and 17H01103.
Y.H. is supported by the Jet Propulsion Laboratory, California Institute of Technology, under a contract with the National Aeronautics and Space.
H.B.L. is supported by the Ministry of Science and Technology (MoST) of Taiwan (grant Nos. 108-2112-M-001-002-MY3).
This paper makes use of the following ALMA data: ADS/JAO.ALMA\#2017.1.00286.S. ALMA is a partnership of ESO (representing its member states), NSF (U.S.), and NINS (Japan), together with NRC (Canada), NSC and ASIAA (Taiwan), and KASI (Republic of Korea), in cooperation with the Republic of Chile.
The joint ALMA observatory is operated by ESO, auI/NRAO, and NAOJ.
Data analysis was carried out on the Multi-wavelength Data Analysis System operated by the Astronomy Data Center (ADC), National Astronomical Observatory of Japan.
Radiative transfer simulations were carried out on analysis servers at Center for Computational Astrophysics, National Astronomical Observatory of Japan.

\software{RADMC-3D \citep{Dullemond_Dominik2005}, CASA v5.6 \citep{McMullin2007}, vis\_sample \url{https://github.com/AstroChem/vis_sample}, emcee \citep{emcee}, Matplotlib \citep[\url{http://matplotlib.org}]{Matplotlib}, NumPy \citep[\url{http://www.numpy.org}]{NumPy}}

\appendix
\section{Statistics of visibility data} \label{sec:vis_stat}
\begin{figure*}
\centering
\resizebox{0.98\textwidth}{!}{\includegraphics{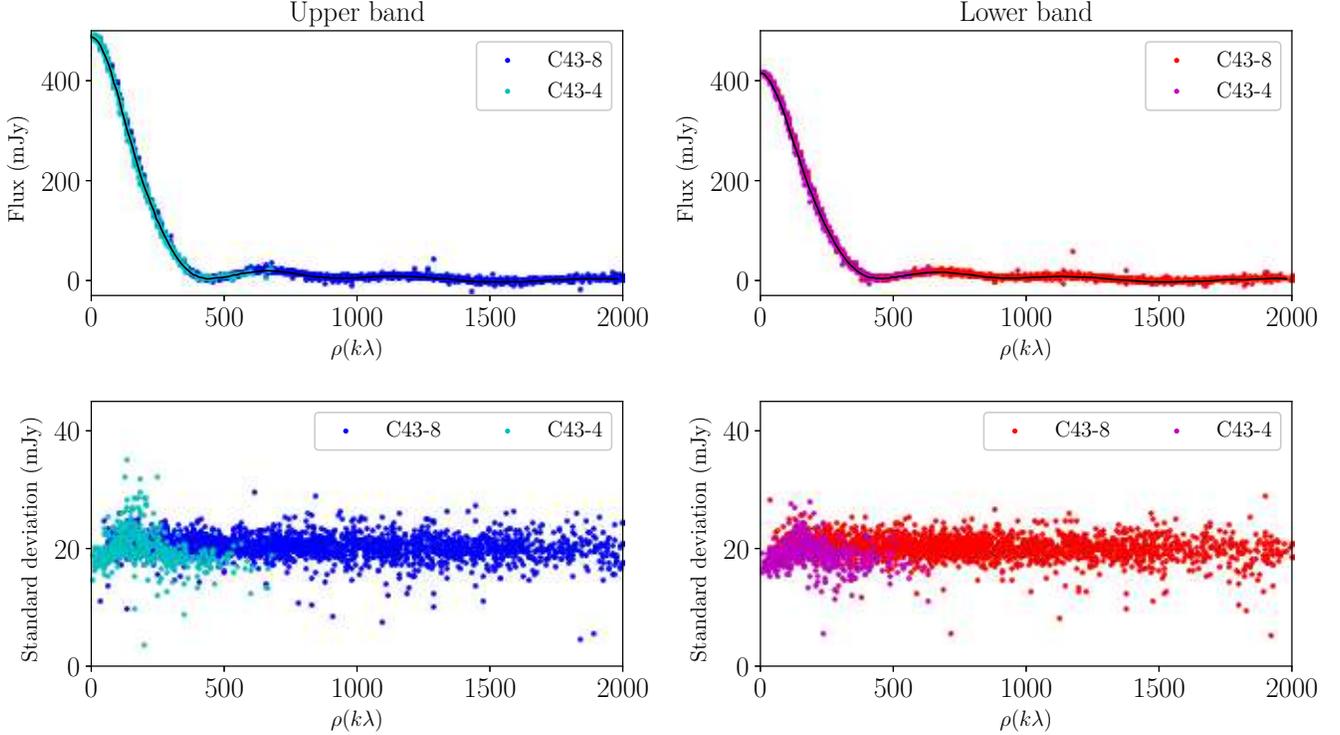}}
\caption{
({\it Upper}) Real parts of the visibility of the C43-4 (compact array configuration) and C43-8 (sparse array configuration) data in the upper and lower bands.
The visibilities are averaged within the bin which the radial width is the same as that described in Section~\ref{sec:obs} and the azimuthal width is $0.2\pi$.
The thin solid lines indicates the visibilities shown in Figure~\ref{fig:vis_diffband}.
({\it Lower}) Standard deviation of the data within the bin.
\label{fig:vis_real_spws}}
\end{figure*}
In this appendix, we show the statistics of the visibility data.
In the upper panel of Figure~\ref{fig:vis_real_spws}, we shows the real parts of the visibility for the C43-4 (compact configuration) and C43-8 (sparse configuration) data in the upper and lower bands, separately.
The visibility in the figure is averaged the bin of the same $uv$-distance width as Figure~\ref{fig:vis_diffband} but with the azimuthal width of $0.2\pi$, instead of whole $2\pi$ in Figure~\ref{fig:vis_diffband}.
Hence, the figure enable us to see the scatter of data in the azimuthal direction.
For reference, we plot the visibility shown in Figure~\ref{fig:vis_diffband}.
In the lower panel of Figure~\ref{fig:vis_real_spws}, we shows the standard deviation of the data within the bin.
The visibilities of the C43-4 and C43-8 data are similar to each other.
The standard deviations are similar in all the data, namely $\sigma \simeq 20\mbox{ mJy}$.
However, around $\rho=200k\lambda$, there are some points with larger standard deviations in the C43-4 data (especially at the upper band data).
Because of this data scatter, the value of $\chi^2$ increases around $\rho=200k\lambda$ when the visibility combined by the short and long baseline data for the MCMC fitting.

\begin{figure*}
\centering
\resizebox{0.98\textwidth}{!}{\includegraphics{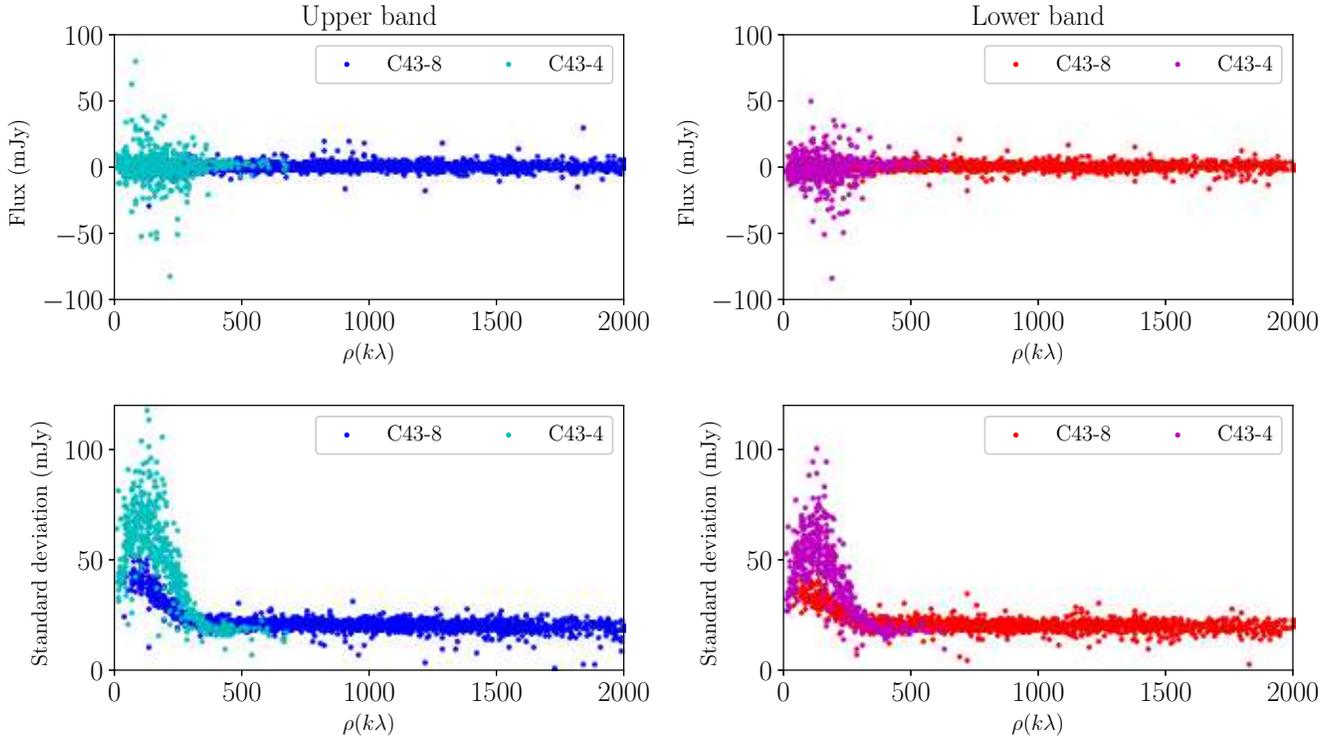}}
\caption{
The same as Figure~\ref{fig:vis_real_spws}, but for the imaginary parts.
\label{fig:vis_imag_spws}}
\end{figure*}
Figure~\ref{fig:vis_imag_spws} is the same as Figure~\ref{fig:vis_real_spws} but for the imaginary part of the visibility.
The standard deviation of the imaginary part is larger than that of the real part in the shorter baseline, namely $\rho \lesssim 200k\lambda$, whereas it is comparable with that of the real parts at long baseline.
Hence, the imaginary part of the visibility shown in Figure~\ref{fig:vis_imag} has a relatively large error at short baseline.

\section{Posteriors of MCMC fitting} \label{sec:mcmc_posterior}
\begin{figure*}
\centering
\resizebox{0.98\textwidth}{!}{\includegraphics{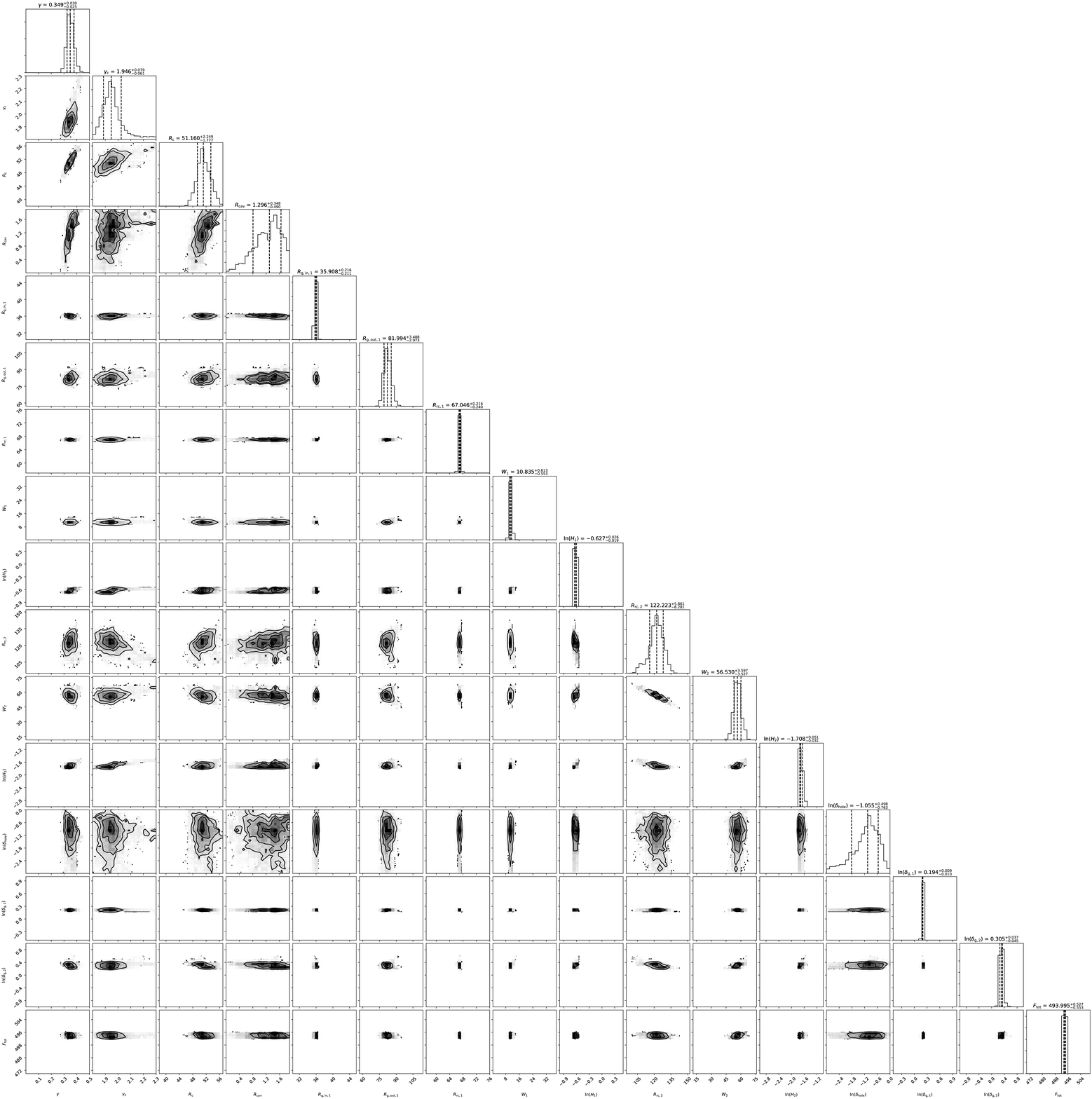}}
\caption{
Posterior of the MCMC fitting for the upper band data.
\label{fig:mcmc_posterior_spw0}}
\end{figure*}

\begin{figure*}
\centering
\resizebox{0.98\textwidth}{!}{\includegraphics{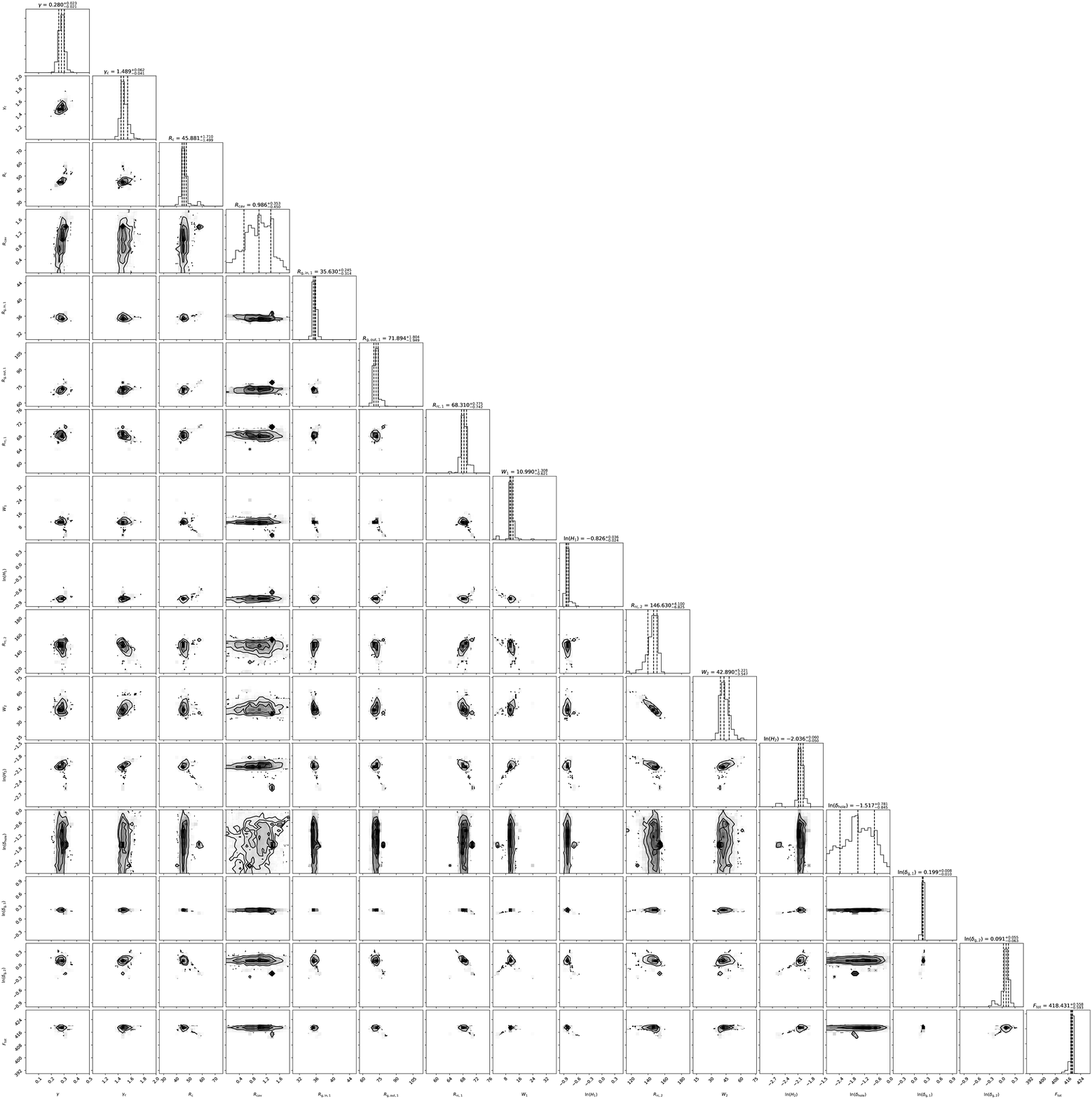}}
\caption{
Posterior of the MCMC fitting for the lower band data.
\label{fig:mcmc_posterior_spw1}}
\end{figure*}

The posterior of the MCMC fitting for the upper hand data is shown in Figure~\ref{fig:mcmc_posterior_spw0} and that for the lower band data is shown in Figure~\ref{fig:mcmc_posterior_spw1}.

\section{Spectral index map with a power-law size distribution for large dust grains} \label{sec:rt_powd}
In Section~\ref{sec:rt}, we carried out radiative transfer simulations with large grains which has a Gaussian size distribution around the specific radius of the dust grains.
Here, we show the results of radiative transfer simulations with large grains which has a power-law size distribution like that of the small grains, and investigate the dependence of the dust size distribution on the spectral index map.

As in the cases described in Section~\ref{sec:rt}, we consider two-kind of dust grains: one represents small grains, other is large grains.
The size distribution of the small dust is the same as that described in Section~\ref{sec:rt}.
For the large grains, the number density of the dust grains is proportional to $s^{p}$, and we adopted $p=-3.5$ and $-2.5$ cases.
The minimum size of the large grains is $0.1$~mm, and we consider three maximum sizes of the large grains, namely, $s_{\rm max}=$ 1~mm and 0.5~mm. 
The stellar luminosity is $22L_{\odot}$ and $\Sigma_0=1.5\mbox{ g/cm}^2$, and other parameters are the same these described in Section~\ref{sec:rt}.
 

\begin{figure}
\centering
\resizebox{0.49\textwidth}{!}{\includegraphics{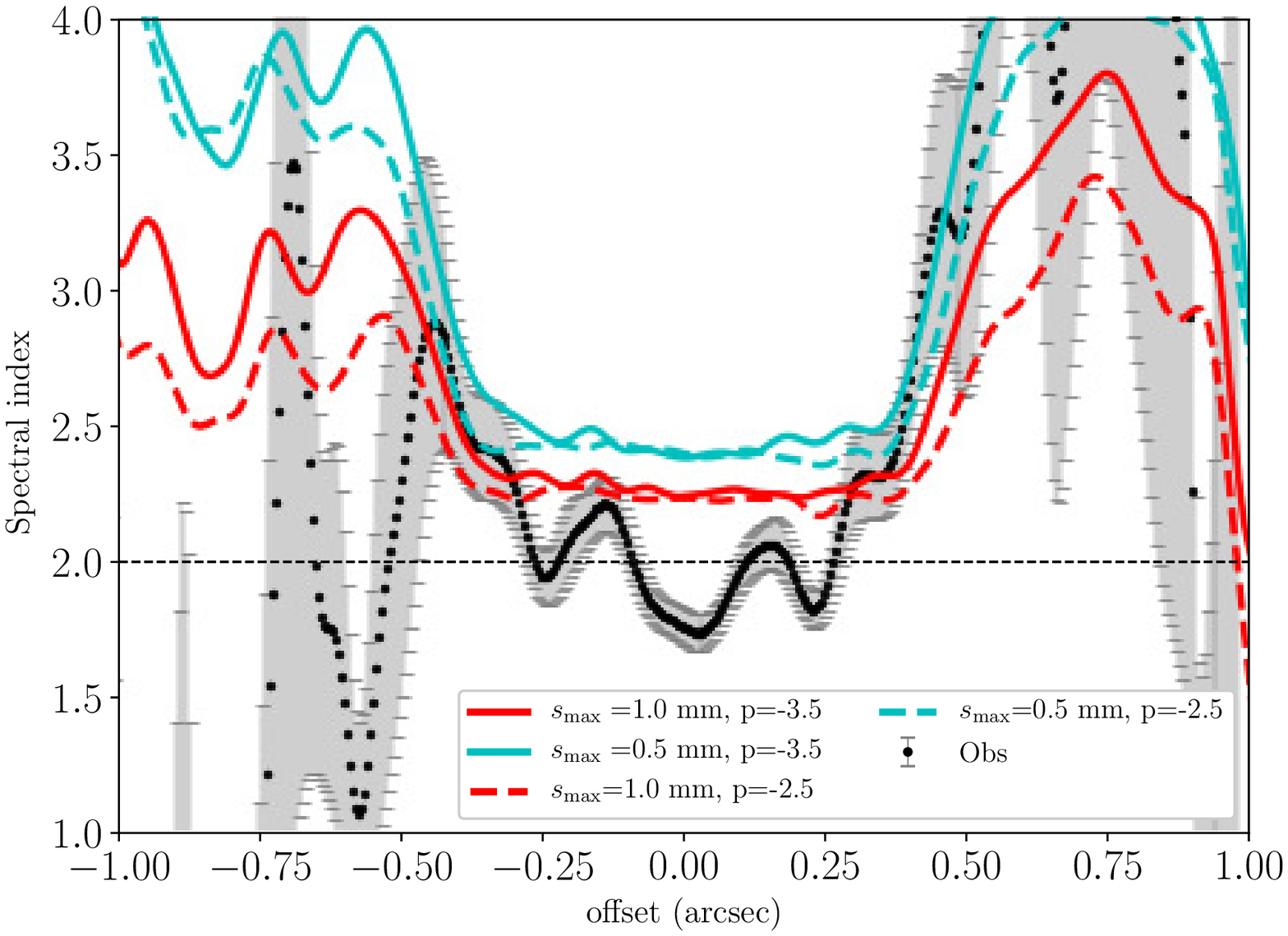}}
\caption{
The same as Figure~\ref{fig:spectral_index_slice_mcrt}, but for the power-law sized large grains, and in the case of $p=-3.5$ (solid lines) and the case with $p=-2.5$ (dashed lines).
\label{fig:alpha_slice_mcrt_powd}}
\end{figure}
Figure~\ref{fig:alpha_slice_mcrt_powd} shows the spectral index along the major axis for the cases with $p=-3.5$ and $p=-2.5$.
In all the cases, the spectral index is about $2$ (but slightly larger than 2), and it increases in outer region.

\section{Spectral index map for ALMA bands} \label{sec:alpha_diff_bands}
For future observations, in this appendix, we present a few examples of the spectral index map given by radiative transfer simulations.
As in the previous section, $L_{\odot}=22L_{\odot}$ and $\Sigma_0=1.5\mbox{ g/cm}^2$, and other parameters are the same these described in Section~\ref{sec:rt}.
As different from these shown in Section~\ref{sec:rt}, the spectral index presented in this section is not calculated through the \verb#tclean# task.
The spectral index is calculated by just the differences of the fluxes convoluted with a Gaussian filter with $0.1$~arcsec standard deviation.
We calculated the spectral index as the difference of the fluxes among ALMA band~3, band~6, band~7, and band~9.

\begin{figure}
\centering
\resizebox{0.49\textwidth}{!}{\includegraphics{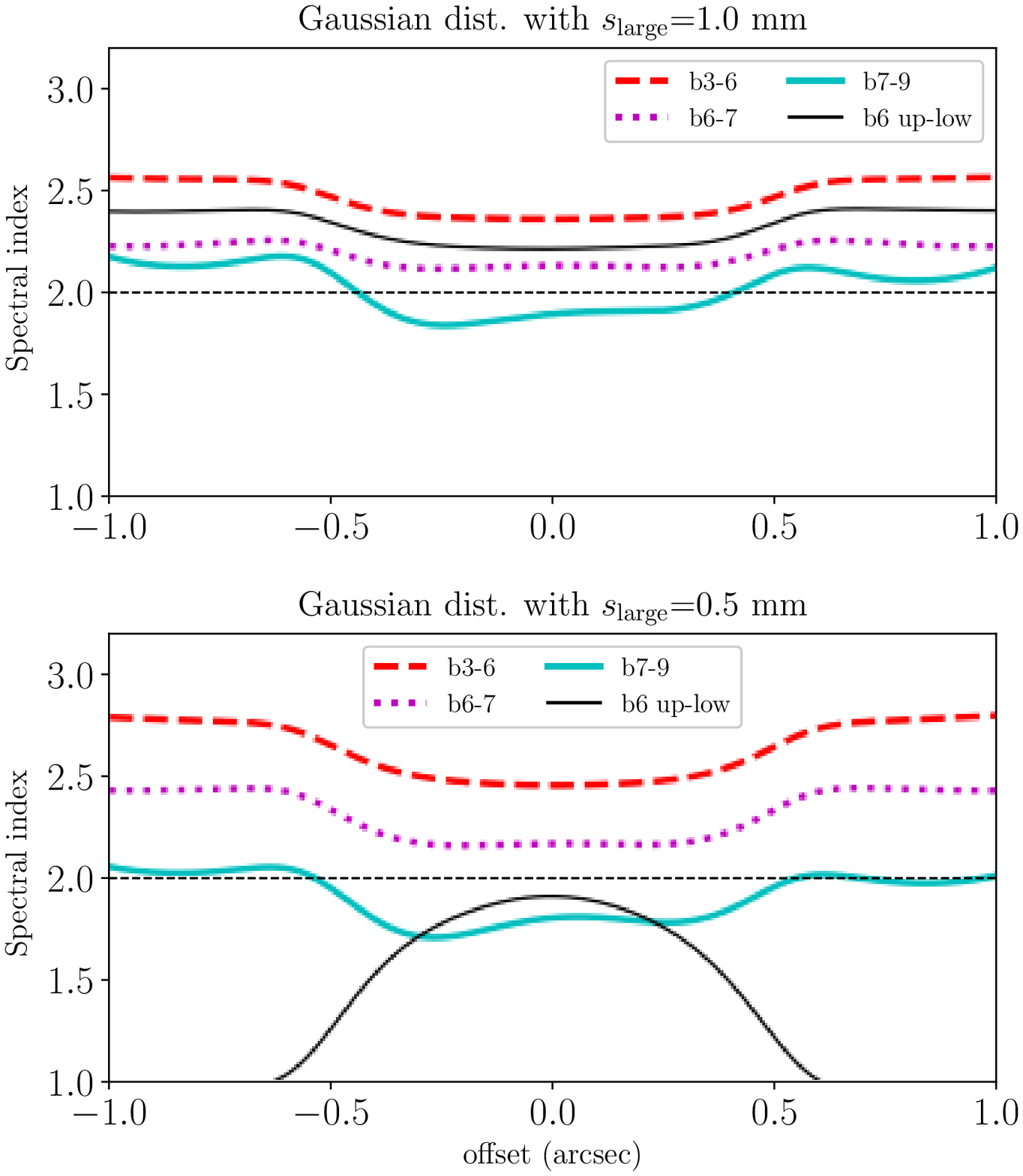}}
\caption{
The spectral index distributions along the major axis in the cases that the large dust grains have a Gaussian size distribution with $\slarge=1$~mm (top), $0.5$~mm (bottom).
The dashed, dotted, and solid lines indicate the spectral indices calculated from the fluxes at band 3 (3.1~mm) and at band~6 (1.3~mm), the fluxes at band~6 and at band~7 (0.87~mm), and the fluxes at band~7 and band~9 (0.45~mm), respectively.
The thin solid line indicates the spectral index given by the upper and lower bands of band~6 as the same as presented in the main text, for reference.
\label{fig:alpha_gaussian}}
\end{figure}
\begin{figure}
\centering
\resizebox{0.49\textwidth}{!}{\includegraphics{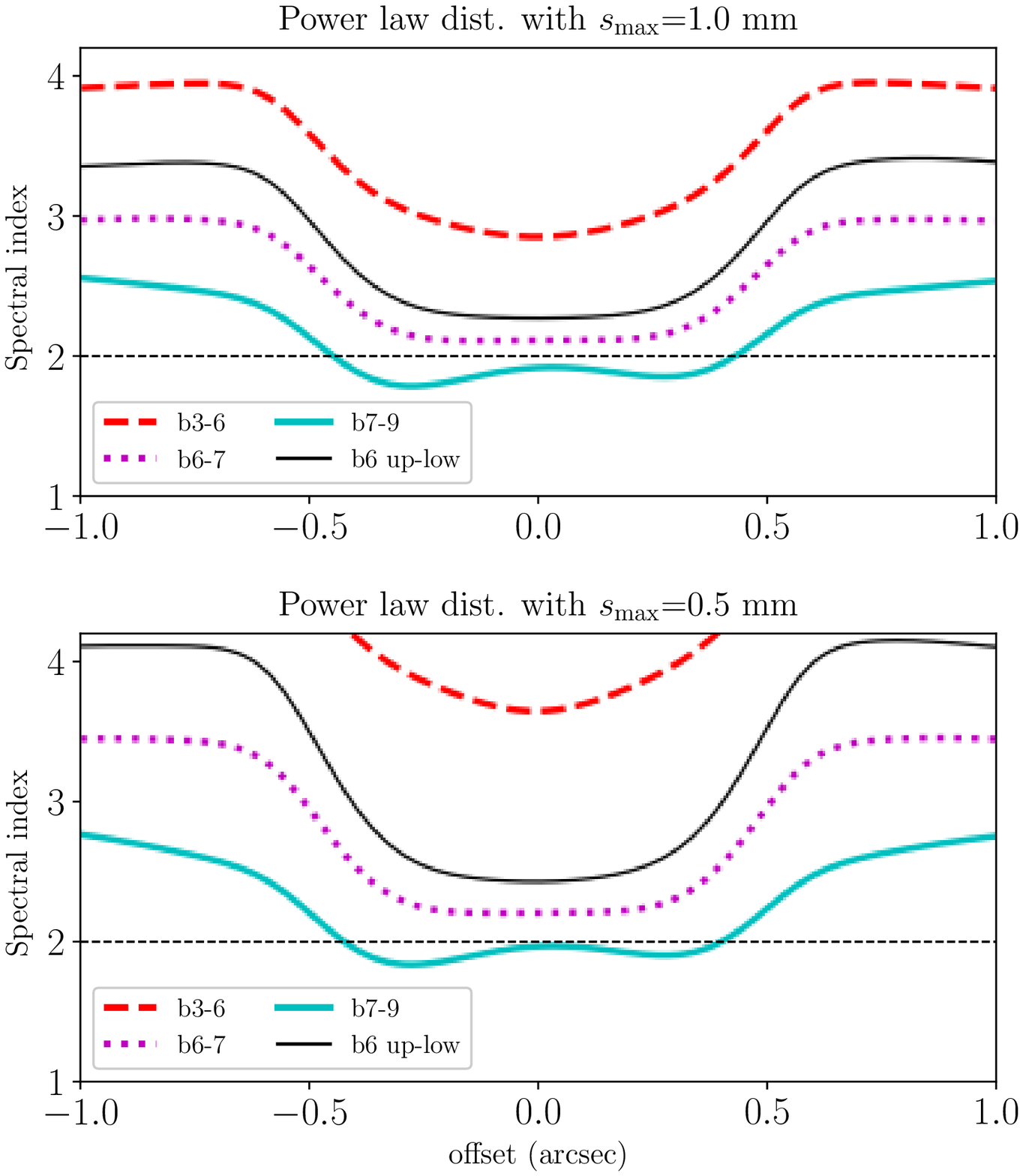}}
\caption{
The same as Figure~\ref{fig:alpha_gaussian}, but the large grains have a power-law size distribution.
\label{fig:alpha_powlaw}}
\end{figure}
Figure~\ref{fig:alpha_gaussian} shows the spectral index along the major axis, when the large grains has a Gaussian (log-normal) distribution (for the detail, see Section~\ref{sec:rt}).
The spectral index depends on the choice of the bands used.
The spectral index calculated from bands with longer wavelengths becomes larger , as the opacity is smaller and it is optically thin.
For instance, the spectral index calculated from the band~3 and band~6 fluxes are larger than the spectral index calculated from other pairs of bands.
The spectral index calculated from the band~7 and band~9 is below 2 around the center, regardless of $\slarge$.
It is worth pointing out that the spectral index calculated from band 6 and band 7 significantly different from that calculated from the upper and lower bands of band~6, in the cases with $\slarge = 0.5$~mm.
Figure~\ref{fig:alpha_powlaw} shows the same as that shown in Figure~\ref{fig:alpha_powlaw}, but for the cases that the large dust grains has a power-law distribution of $\propto s^{-3.5}$.
The distributions of the spectral index calculated by band~6 and 7 is similar to that in the case with the Gaussian distribution shown in Figure~\ref{fig:alpha_gaussian}, which is below 2 around the center.
On the other hand, the spectral index calculated from the band~3 and 6 is larger than that in the case with the power-law distributions, as compared with that in the cases with the Gaussian distribution.
We may be able to constrain the dust size distribution from the difference of the spectral indexes calculated from the different pair of the bands.


\end{document}